\documentclass[twocolumn]{aastex61}

\newcommand\pampelmuse{\texttt{PampelMuse}}

\received{\today}
\revised{}
\accepted{}

\submitjournal{AJ}

\shorttitle{Westerlund 2 observed with MUSE}
\shortauthors{Zeidler et al.}

\begin{document}

\title{The young massive star cluster Westerlund 2 observed with MUSE. \\
	I. First results on the cluster internal motion from stellar radial velocities}

\correspondingauthor{Peter Zeidler}
\email{zeidler@stsci.edu}

\author[0000-0002-6091-7924]{Peter Zeidler}
\affil{Space Telescope Science Institute , 3700 San Martin Drive, Baltimore, MD 21218, USA}

\author{Elena Sabbi}
\affil{Space Telescope Science Institute , 3700 San Martin Drive, Baltimore, MD 21218, USA}

\author{Antonella Nota}
\affil{Space Telescope Science Institute , 3700 San Martin Drive, Baltimore, MD 21218, USA}
\affil{ESA, SRE Operations Devision, Spain}

\author{Anna Pasquali}
\affil{Astronomisches Rechen-Institut, Zentrum f\"ur Astronomie der Universit\"at Heidelberg, M\"onchhofstra{\ss}e 12--14, D-69120 Heidelberg, Germany} 

\author{Eva K. Grebel}
\affil{Astronomisches Rechen-Institut, Zentrum f\"ur Astronomie der Universit\"at Heidelberg, M\"onchhofstra{\ss}e 12--14, D-69120 Heidelberg, Germany} 

\author{Anna Faye McLeod}
\affil{School of Physical and Chemical Sciences, University of Canterbury, New Zealand}

\author{Sebastian Kamann}
\affil{Astrophysics Research Institute, Liverpool John Moores University, 146 Brownlow Hill, Liverpool L3 5RF, United Kingdom}
\affil{Institute for Astrophysics, Georg-August-University, Friedrich-Hund-Platz 1, D-37077 G\"ottingen, Germany} 

\author{Monica Tosi}
\affil{INAF - Osservatorio di Astrofisica e Scienza dello Spazio, Via Gobetti 93/3, I-40129, Bologna, Italy}

\author{Michele Cignoni}
\affil{Dipartimento di Fisica, Universita' di Pisa, Largo Bruno Pontecorvo, 3, 56127 Pisa, Italy}
\affil{INFN, Sezione di Pisa, Largo Pontecorvo 3, 56127 Pisa, Italy}

\author{Suzanne Ramsay}
\affil{ESO/European Southern Observatory, Karl-Schwarzschild-Stra{\ss}e 2, 85748, Garching bei M\"unchen, Germany}


%
%
%

%

\begin{abstract}
Westerlund 2 (Wd2) is the central ionizing star cluster of the \ion{H}{2} region RCW~49 and the second most massive young star cluster (${\rm M} = (3.6 \pm 0.3)\times 10^4\,{\rm M}_\odot$) in the Milky Way. Its young age ($\sim2\,$Myr) and close proximity to the Sun ($\sim 4\,$kpc) makes it a perfect target to study stars emerging from their parental gas cloud, the large number of OB-stars and their feedback onto the gas, and the gas dynamics. We combine high-resolution multi-band photometry obtained in the optical and near-infrared with the \textit{Hubble} Space Telescope (HST), and VLT/MUSE integral field spectroscopy to study the gas, the stars, and their interactions, simultaneously. In this paper we focus on a small, $64\times64\,{\rm arcsec}^2$ region North of the main cluster center, which we call the Northern Bubble (NB), a circular cavity carved into the gas of the cluster region. Using MUSE data, we determined the spectral types of 17 stars in the NB from G9III to O7.5. With the estimation of these spectral types we add 2 O and 5 B-type stars to the previously published census of 37 OB-stars in Wd2. To measure radial velocities we extracted 72 stellar spectra throughout Wd2, including the 17 of the NB, and show that the cluster member stars follow a bimodal velocity distribution centered around $(8.10 \pm 1.53)\,{\rm km}\,{\rm s}^{-1}$ and $(25.41 \pm 1.57)\,{\rm km}\,{\rm s}^{-1}$ with a dispersion of $(4.52 \pm 1.78)\,{\rm km}\,{\rm s}^{-1}$ and $(3.46 \pm 1.29)\,{\rm km}\,{\rm s}^{-1}$, respectively. These are in agreement with CO($J=1$--2) studies of RCW~49 leaving cloud-cloud collision as a viable option for the formation scenario of Wd2. The bimodal distribution is also detected in the Gaia DR2 proper motions.

\end{abstract}


\section{Introduction}
\label{sec:introduction}

The young massive stars cluster (YMC) Westerlund 2 \citep[Wd2,][]{Westerlund_61} is the central ionizing cluster of the \ion{H}{2} region RCW49 \citep{Rodgers_60} and is the second most massive YMC in the Milky Way \citep[MW, total stellar mass: $(3.6\pm0.3) \times 10^4 {\rm M}_\odot$,][]{Zeidler_17} located in the Sagittarius spiral arm $\left( \alpha,\delta \right) = \left(10^{\rm h }23^{\rm m}58^{\rm s},-57^\circ45'49''\right)$(J2000) (l,b)=$\left(284^\circ.3, -0^\circ.34\right)$. With its young age of $1-2$~Myr, its close proximity to the Sun \citep[4.16~kpc,][]{Vargas_Alvarez_13,Zeidler_15}, and its high-mass stellar content \citep[37 spectroscopically identified OB-type stars,][]{Moffat_91,Vargas_Alvarez_13} it is a perfect testbed to study the early evolution and feedback of YMCs. \citet{Moffat_91} also suggests that Wd2 contains more than 80 O-type stars.

We have studied Wd2 photometrically using \textit{Hubble} Space Telescope (HST) multi-band data (ID: 13038, PI: A. Nota) obtained in the optical and infrared \citep{Zeidler_15,Zeidler_16b,Zeidler_17} with the Advanced Camera for Surveys \citep[ACS,][]{ACS} and the infrared channel of the Wide Field Camera 3 \citep[WFC3/IR,][]{WFC3}. In addition to the age and distance estimate, we confirmed the finding by \citet{Hur_15} that Wd2 consists of two sub-clumps \citep{Zeidler_15}, the main cluster (MC) and the northern clump (NC). Both clumps appear to be coeval. We derived a stellar mass function (MF) of the whole cluster area, as well as different sub regions (using elliptical annuli centered on the cluster center), and we showed that the high-mass slope of the MF is $\Gamma=-1.53 \pm 0.05$, steeper than a \citet{Salpeter_55} slope of $\Gamma = -1.35$. This is quite common among YMCs, e.g.: $\Gamma=-1.44^{+0.56}_{-0.08}$ for Westerlund~1 \citep{Gennaro_11} or $\Gamma=-1.87\pm 0.41$ for NGC~346 \citep{Sabbi_08}. A study of the evolution of the MF slope with increasing radii from the cluster center revealed that Wd2 is highly mass segregated, and given the young age, the mass segregation is likely primordial.

Combining HST photometric wide and narrow-band filters, 240 bona-fide pre-main sequence (PMS) H$\alpha$ excess emitters were identified \citep{Zeidler_16b} indicating still ongoing mass accretion on the host stars. The analysis of the mass accretion hinted at an increase of the mass accretion rate with distance to the luminous OB stars. This suggests that the high amount of FUV flux radiated by the OB stars leads to a faster disk dispersal in close proximity to massive stars.

Studying Galactic YMCs spectroscopically has traditionally been challenging. Slit and fiber spectrographs only allow a very limited number of stars to be observed with a reasonable allocation of telescope time. This has changed in the past decade, when the development of integral field units (IFUs) has made major progress. Using the Multi Unit Spectrographic Explorer \cite[MUSE,][]{Bacon_10}, mounted in the Nasmyth focus of UT4 at the Very Large Telescope (VLT) allows us, for the first time, to efficiently map Galactic star clusters spectroscopically. MUSE has a field-of-view (FOV) of $1~\rm{arcmin}^2$ at a spatial sampling of $0.2~\rm{arcsec~px}^{-1}$ and a resolving power of $R \approx 2000$--4000 at optical wavelengths between $4650\rm{\AA}$ and $9300\rm{\AA}$. This gives a total number of spectral pixels (spaxels) of $\sim96000$ and $3800$ wavelength bins with $\Delta \lambda =1.25\rm{\AA}$. In the past, MUSE has been proven to be an excellent instrument to reveal motions, abundances, and 3D structures of gas and molecular clouds \citep[e.g., the Pillars of Creation in M16 and the central Orion nebula,][]{McLeod_15,McLeod_16}. In addition, MUSE has been used to measure the stellar radial velocities and velocity dispersions of, e.g., the globular cluster NGC 6397 \citep{Kamann_16} or the ultra-faint stellar system Crater/Laevens I \citep{Voggel_16}.

In this paper we show that, in combination with high-resolution photometry from HST, MUSE is a powerful instrument to study both the stellar and gas content in YMCs and gives us the opportunity to estimate the velocity dispersion with a high enough accuracy to determine whether Wd2 is massive enough to be long-lived. If Wd2 is massive enough to survive the sudden changes in the gravitational potential, as soon as the massive OB star population explodes, this cluster may provide new insight to answer if YMCs are possible progenitors to globular clusters \citep[e.g.,][]{Kruijssen_15}. The stellar velocities shed light on the formation and history of Wd2 to see which cluster-formation theory applies: monolithic \citep[e.g.,][]{Lada_84a,Banerjee_15a} or hierarchical \citep[e.g.,][]{Parker_14} cluster formation, cloud-cloud collision \citep[e.g.,][]{Nigra_08,Cignoni_09,Fukui_14}, or even a combination of all three together.

This paper is structured as follows: In Sect.~\ref{sec:NC} we give an overview of the studied region. In Sect.~\ref{sec:data} we introduce our dataset and the data reduction. In Sect.~\ref{sec:extracted_spectra} we analyze the extracted spectra and describe the spectral typing. In Sect.~\ref{sec:stellar_velocities} we describe the technique for measuring the radial velocities. In Sect.~\ref{sec:results} we discuss our findings and compare results, while in Sect.\ref{sec:conclusions}  we summarize our findings and provide a future outlook on this project.

\section{The Northern Bubble}
\label{sec:NC}

\begin{figure*}[htb]
	\plotone{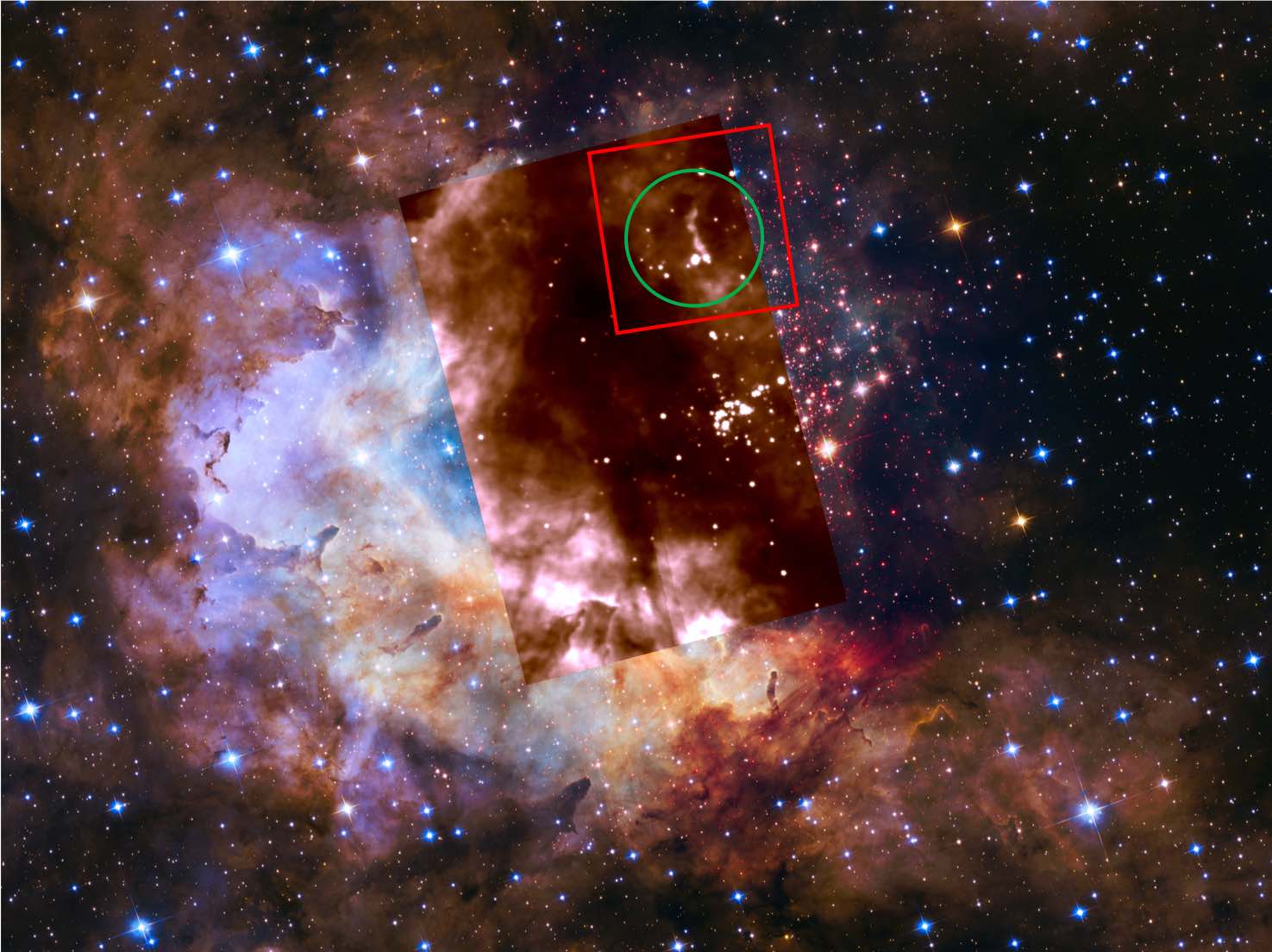}
	\caption{Color composite image of the HST ACS and WFC3/IR data of Wd2, including the $F125W$ (red), $F814W$ (green), and $F555$W (blue) filters. We present the mosaic of the 6 observed MUSE cubes (short exposures) as an inlay. The red box marks the $64'' \times 64''$ NB region, while the green circle marks the NB. North is up, east to the left. HST image Credit: NASA, ESA, the Hubble Heritage Team (STScI/AURA), A. Nota (ESA/STScI), and the Westerlund 2 Science Team.}
	\label{fig:mosaic}
\end{figure*}

\begin{figure}[htb]
	\plotone{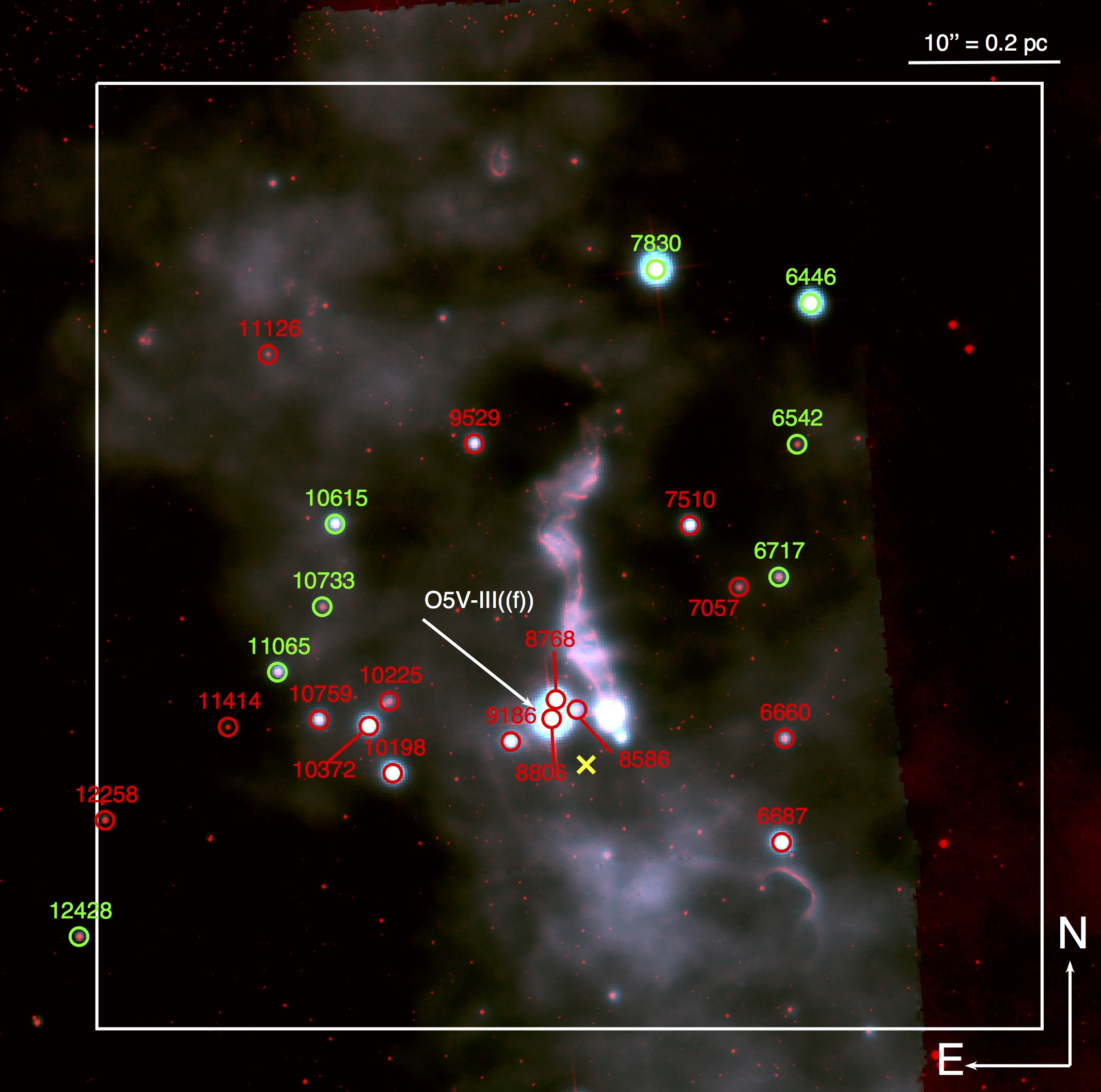}
	\caption{A color-composite image, representing emission in the HST H$\alpha$ ($F658N$ filter, red), \ion{N}{2} 6583\AA~(green), and \ion{O}{3} 5007\AA~(blue) filters, covering the $64'' \times 64''$ field examined in this work (red square, see also Fig.~\ref{fig:mosaic}). The HST H$\alpha$ is used to see small scale structure, below the resolution limit of MUSE, as well as faint stars. The yellow cross marks the center of the NC \citep{Zeidler_17}. The stars analyzed in this work are marked in red for NB members and in green for field stars. The numbers show the catalog identifier (see also Tab.~\ref{tab:spec_stars}).}
	\label{fig:NB}
\end{figure}

\citet{Hur_15}, \citet{Zeidler_15}, and \citet{Zeidler_17} pointed out that Wd2 is built up of two clumps, the Main Cluster (MC) and the Northern Clump (NC). The NC is located $\sim1$pc North of the MC and has a photometric mass of $\left(3 \pm 0.3 \right) \cdot 10^3 {\rm M}_\odot$. In this paper we mainly focus on a sub-region of our dataset covering the area around the NC forming a cavity in the gas distribution of the \ion{H}{2} region (see Fig.~\ref{fig:mosaic} and Fig.~\ref{fig:NB}).  From now on we call this region the Northern Bubble (NB). In the center of this cavity lies a pillar or jet-like gas structure, which we refer to as "the Sock". The NB was first mentioned by \citet{Vargas_Alvarez_13} as a ring-like structure. They suggested it to be the boundary of the \ion{H}{2}-region surrounding a luminous O5V-III((f)) star \citep[Fig~\ref{fig:NB} and ][]{Rauw_07}. By looking at \textit{Spitzer} images of this region, \citet{Vargas_Alvarez_13} concluded that this ring "is present in all IRAC bands but best seen at $[8.0]$, consistent with PAH emission from a photodissociation region". This structure is thus representative of the physical interplay between massive stars and their surrounding ISM. The fact that it is characterized by a low stellar density (thus crowding effects are a lesser issue) makes it a perfect testbed for developing analysis routines, later applicable to the full dataset (see Sect.~\ref{sec:data}).

\section{Data reduction and source extraction}
\label{sec:data}

\subsection{The observations}
We observed Wd2 with MUSE in extended mode during the ESO period 97 (Program ID: 097.C-0044(A), PI: P.~Zeidler). During the night of 2016, June 02/03, we acquired 6 pointings with 3 dither positions each for a total exposure time of $660~\rm{s}$. The dither pattern follows a 90 and 180 degrees rotation strategy to minimize detector defects and impurities.  The observations were obtained in two observation blocks (OBs): 1323877 and 1323880. The seeing ranges from $0.61'' - 1.1''$ with an airmass of $1.21 - 1.28$ and $0.49'' - 0.78''$ with an airmass of $1.29 - 1.43$, respectively. A detailed overview of the data is presented in \citet{Zeidler_18b}. A second proposal has been approved and observed covering the whole Wd2 region, including some of the surrounding gas. In total we have obtained 15 pointings, including deep ($3 \times 3600\,\rm{s}$) exposures to observe the PMS down to 1--2\,M$_\odot$. We used the standard reduction pipeline (v.2.0.1) provided by ESO\footnote{\url{https://www.eso.org/sci/software/pipelines/muse/}} \citep{Weilbacher_12,Weilbacher_14} to reduce the data and combine the dither pattern into 6 data cubes (one for each pointing). This pipeline is based on the ESO Reflex environment (ESORex) for automated data reduction work flows for astronomy \citep{Freudling_13}. The datacubes are wavelength calibrated and the radial velocity of the telescope is corrected to the barycenter of the Solar System. We visually inspected the datacubes to determine that the different dither positions were properly aligned and that the data reduction was a success. The mosaic of the six data cubes is shown in Fig.~\ref{fig:mosaic} as an inlay in the HST color-composite image of the Wd2 region. The spatial sampling is $0.2''\,\rm{px}^{-1}$ with a spectral sampling of 1.25\,\AA, the resolution is 2.4\,\AA ($\rm{R} \approx 2000$--4000 from blue to red).

\subsection{The source extraction}

Analyzing MUSE data obtained in crowded regions, such as YMCs, is challenging. We need to detect and mask stars in order to investigate the gas \citep[e.g.,][]{McLeod_15} but for the stellar source extraction, a wavelength-dependent point spread function (PSF) and wavelength-dependent background have to be taken into account. This is done with ''PampleMuse'', a python package developed by Sebastian Kamann \citep{Kamann_13} that uses a deep (at least 2~mag deeper than the detection limit of the MUSE observations), high-resolution photometric catalog to perform PSF spectrophotometry on the pipeline-reduced MUSE data cubes. In their study of simulated MUSE data of a crowded field, \citet{Kamann_13} showed that it is possible to extract  $\sim5000$ useful stellar spectra per arcmin$^2$. Here we give a short overview of the source extraction procedure. For a detailed description, see \citet{Zeidler_18b}:

\begin{itemize}
	\item [1)] As first step, PampelMuse selects isolated (low crowding) bright sources to perform PSF spectrophotometry. This defines a wavelength-dependent PSF and corrects for a spatial offset or rotation with respect to the reference catalog and for wavelength-dependent variations in the positions.
	\item [2)] The wavelength-dependent PSF profile and coordinate corrections are then used for all of the stellar sources down to a signal-to-noise (S/N) based brightness limit. With this method sources separated less than their full-width half maximum (FWHM) can still be extracted.
	\item [3)] The background around each star is estimated by subdividing the FOV in sub-regions.
	\item [4)] The final products are background-subtracted spectra of the stars in the FOV.
	
\end{itemize}

From our parent sample that will be presented in \citet{Zeidler_18b} and that contains all stars covered by the short and long exposures down to a ${\rm S/N}\footnote{The S/N of the stellar spectra is calculated over full spectral range} \approx 5$, we selected all stars in the NB with a ${\rm S/N} \ge 20$. In addition, we selected 55 stars with ${\rm S/N} \ge 20$ from the other parts of the MUSE mosaic to study the cluster dynamics.

\section{The MUSE spectra}
\label{sec:extracted_spectra}
The background subtraction described in Sect.~\ref{sec:data} works well for stars that are located in areas with a low background variability. If for example, a star is located in front of a gas ridge, the background emission coming from the gas is highly different on one side of the star compared to the other, which leads to a gradient in the background distribution. This becomes especially apparent at wavelengths where stars have absorption lines and the gas has strong emission lines (e.g,. hydrogen and helium lines), leading to either an over or under-subtraction of the background. We inspected the PampelMuse extracted spectra by eye and, as an indicator, we used the [\ion{N}{2}]$\lambda\lambda6549,6583$ lines to evaluate the quality of the background subtraction. These two lines are purely nebular lines and should totally disappear in the stellar spectra. In addition, we used strong lines, especially H$\alpha$ and H$\beta$, which often show negative fluxes in case of a background over-subtraction. With this method we selected those stars, for which we have spectra with a well-subtracted background.

\subsection{Normalization and line identification}
To normalize and rectify the extracted spectra we use the spectral fitting package \texttt{pyspeckit}\footnote{\url{https://github.com/pyspeckit/pyspeckit}, Authors: Adam Ginsburg, Jordan Mirocha, pyspeckit@gmail.com}. This python based routine allows us to fit spectral lines (absorption and emission) together with a continuum in an iterative process to determine an optimal solution. Due to the wavelength range of more than $4000~\rm{\AA}$ we split the wavelength range into 22 blocks to find a reliable continuum solution using low-degree polynomials.

To identify stellar spectral features, we used several sources in the literature, such as \citet[][and references therein]{Gray_09,Kaler_11,Rauw_11,Sota_11,Sota_14}. To obtain the exact rest wavelengths of the absorption lines we used the ''NIST Atomic Spectra Bibliographic Databases''\footnote{\url{https://physics.nist.gov/PhysRefData/ASD/lines_form.html}}. For the identification of diffuse interstellar bands (DIBs) we used the ''DIB Database''\footnote{\url{http://dibdata.org}} and specifically the studies of the stellar spectrum of HD183143 by \citet{Hobbs_09}.

The spectra of early-type stars (mostly O and B) are typically dominated by neutral and ionized helium lines, while common features of late-type stars are neutral and ionized metals and the pronounced \ion{Ca}{2}-triplet, typical of cooler atmospheres.

\subsection{Spectral Classification}
\label{sec:spec_class}
To estimate the spectral type of each of the stars in our sample, we performed a multi-step classification. We sorted the stars in four major categories:

\begin{itemize}
	\item[1)] Stars showing  \ion{He}{1} and \ion{He}{2} absorption lines are classified as O-type stars.
	\item[2)] Stars showing \ion{He}{1} but no \ion{He}{2} are classified as B-type stars.
	\item[3)] Stars that show no helium lines but strong hydrogen lines, and a few ionized metals are classified as A-type stars.
	\item[4)] Stars showing the prominent \ion{Ca}{2}-triplet and ionized and neutral metals are classified as A9 and later.
\end{itemize}

\subsubsection{The O stars}
\label{sec:O_stars}

The classification of O-type stars is usually done in the ultraviolet \citep[$\lambda \le 4600{\rm \AA}$, e.g., ][]{Walborn_90}. Since the MUSE spectra cover optical and NIR wavelengths, we used the empirical work of \citet{Kobulnicky_12}. They used the equivalent width (EW) ratio of \ion{He}{2}$\lambda5411$ over \ion{He}{1}$\lambda5876$ and fitted it against the effective stellar temperature ($T_{\rm eff}$). This relation can be described with a $3^{\rm rd}$-order polynomial as derived by \citet{Vargas_Alvarez_13}:

\begin{eqnarray}
\frac{{\rm EW}\left({\rm HeII}\lambda 5411\right)}{{\rm EW}\left({\rm HeI}\lambda 5876\right)} &=& 1.16208 \times 10^{-12}\,T_{\rm eff}^3 \nonumber \\
- 1.19205 \times 10^{-7}\,T_{\rm eff}^2 &+& 4.22137 \times 10^{-3}\,T_{\rm eff} \nonumber \\
&-& 50.5093.
\label{eq:EW_Otype}
\end{eqnarray}

The EWs are measured using \texttt{pyspeckit}. The transformation from temperature to spectral type is made using the calibrations of parameters for O-type stars by \cite{Martins_05}, specifically:

\begin{eqnarray}
T_{\rm eff}=50838-1995 \times {\rm ST}.
\label{eq:T_ST_transform}
\end{eqnarray}

The individual derived parameters of the found O-stars summarized in Tab.~\ref{tab:O_stars}.

\begin{deluxetable}{rccccc}
	\tablecaption{The O-type stars \label{tab:O_stars}}
	\tabletypesize{\footnotesize}
	\tablewidth{0pt}
	\tablehead{
		\multicolumn{1}{c}{ID} &\multicolumn{1}{c}{EW(\ion{He}{1}) } & \multicolumn{1}{c}{EW(\ion{He}{2}) } &  \multicolumn{1}{c}{$T_{\rm eff}$} &  \multicolumn{1}{c}{spectral type}  & \multicolumn{1}{c}{spec. mass} \\
		\multicolumn{1}{c}{    } &\multicolumn{2}{c}{[\AA]} &   \multicolumn{1}{c}{[K]} &  \multicolumn{1}{c}{ }  & \multicolumn{1}{c}{[M$_\odot$]} 
	}
	\tablecolumns{6}
	\startdata
	10198 & 0.8201 & 0.4664  &  31884  &  O9.5\tablenotemark{a} & 15.55  \\ 
	10372 & 0.8933 & 0.7374  &  33553  &  O8.5 & 18.80  \\ 
	8768  & 0.8352 & 1.0804  & 36155  &  O7.5 & 22.90  \\ 
	7510  &  0.6781 & 0.3010 & 31151  &   B0\tablenotemark{b}  & $<15.55$ \\
	\enddata
	\tablecomments{In this table we present the results and the spectral types of the O-type stars. In column~1 we give the stellar ID \citep[as used in our HST photometric catalog, see ][]{Zeidler_15}. In Columns 2 and 3 the measured EWs are presented, while in Columns 4--6 the resulting $T_{\rm eff}$, the spectral types, and the spectroscopic masses \citep[provided by the models of ][]{Martins_05} are given, respectively. We note here that the spectroscopic masses have an uncertainty 35--50\% \citep{Martins_05}.}
	\tablenotetext{a}{This star was classified as O9.5V by \citet{Vargas_Alvarez_13}}
	\tablenotetext{b}{Although this not an O star, we still show it here due to the presence of a weak \ion{He}{2}$\lambda5411$ line.}
\end{deluxetable}

\subsubsection{The B stars}
\label{sec:Bstars}

A similar relation can be found for B-type stars \citep[see Fig.~4 of][]{Kobulnicky_12} using the EW of \ion{He}{1} and the ratio ${\rm EW}\left({\rm H}\alpha \right)/{\rm EW}\left({\rm HeI}\lambda 5876\right)$. For stars earlier than B3 this method becomes degenerate. Due to the insufficient accuracy of the results of the fitted parameters in \citet{Kobulnicky_12} we used the information given in that paper to perform our own fit with the following results:

\begin{eqnarray}
\frac{{\rm EW}\left({\rm H}\alpha \right)}{{\rm EW}\left({\rm HeI}\lambda 5876\right)} &=& -1.27078  \times 10^{-11}\,T_{\rm eff}^3 \nonumber \\
- 9.17611 \times 10^{-7}\,T_{\rm eff}^2 &-& 2.18026 \times 10^{-2}\,T_{\rm eff} \nonumber \\
&+& 173.99531,
\label{eq:EW_Btype_HeI}
\end{eqnarray}

\begin{eqnarray}
{\rm EW}\left({\rm HeI}\lambda 5876\right) &=& 2.91583 \times 10^{-13}\,T_{\rm eff}^3 \nonumber \\
-2.12788 \times 10^{-8}\,T_{\rm eff}^2 &+& 5.12709 \times 10^{-4}\,T_{\rm eff} \nonumber \\
&-&3.405698.
\label{eq:EW_Btype_HaHeI}
\end{eqnarray}

To estimate the spectral type from the effective temperature we used the results from \citet{Underhill_79}. The relation ST- $\log(T_{\rm eff})$ is remarkably linear, which leads to ${\rm ST}=-8.56 \times \log(T_{\rm eff})+88.02$. An overview of the determined spectral types can be found in Tab.~\ref{tab:B_stars}.

\begin{deluxetable}{rcccc}
	\tablecaption{The B-type stars \label{tab:B_stars}}
	\tabletypesize{\footnotesize}
	\tablewidth{0pt}
	\tablehead{
		\multicolumn{1}{c}{ID} &\multicolumn{1}{c}{EW(\ion{He}{1}) } & \multicolumn{1}{c}{EW(H$\alpha$)} &  \multicolumn{1}{c}{$T_{\rm eff}$} &  \multicolumn{1}{c}{spectral type}  \\
		\multicolumn{1}{c}{    } &\multicolumn{2}{c}{[\AA]} &   \multicolumn{1}{c}{[K]} &  \multicolumn{1}{c}{ } 
	}
	\tablecolumns{5}
	\startdata
	6687 & 0.7560 & 3.5973  &  24880  &  B1.5\tablenotemark{a} \\ 
	9186 & 0.7485 & 4.0304  &  24540  &  B1.5 \\
	7510 & 0.6781  & 3.7377  &  28949   &  B0 \\ 
	9529& 0.5445  & 4.0155  &  16149   &  B5 \\ 
	10759 & 0.6216  & 3.9803 &  17290  &  B4.5 \\
	6660 & 0.6640  & 3.6446 &  18490  &  B4.5\tablenotemark{b} \\
	7057 & 1.0324  & 4.1731 &  18730  &  B4 \\ 
	\enddata
	\tablecomments{In this table we present the spectral typing of the B-type stars. In column~1 we give the stellar ID (as specified in \citet{Zeidler_15}). In Columns 2 and 3 the measured EWs are presented, while in Columns 4 and 5 the resulting $T_{\rm eff}$, and the spectral types are given, respectively.}
	\tablenotetext{a}{This star was classified as B1V by \citet{Rauw_07} and \citet{Rauw_11}.}
	\tablenotetext{b}{Due to low S/N \citet{Vargas_Alvarez_13} suggested a spectral type of late O to early B.}
\end{deluxetable}

\subsubsection{The late-type stars}
\label{sec:late_stars}

For the stars with spectral type A9 or later, the \ion{Ca}{2}-triplet is present. We used the \ion{Ca}{2}-triplet libraries provided by \citet{Cenarro_01} and \citet{Munari_99} together with the Penalized Pixel-Fitting \citep[pPXF][]{Cappellari_04,Cappellari_17} method\footnote{This python based routine cross matches the template and the source spectrum to estimate the optimal kinematic solution.} to find the best-fitting template. We excluded the wavelength range between $8610{\rm \AA}$ and $8634{\rm \AA}$ due to a possible blend of the \ion{Fe}{1}$\lambda8622$ line with a DIB. We chose the best fitting model based on the $\chi^2$ value. For the "traditional" spectral typing with by-eye inspection of the remaining stars we refer to the Appendix~\ref{sec:individual_ST}. Both results, the individual by-eye inspection of the spectral lines and the cross-correlation of the spectra with template libraries are in good agreement, indicating that the automated, scripted method is reliable.

A summary of all the derived spectral types of all stars is given in Column 14 of Tab.~\ref{tab:spec_stars}.

\subsubsection{Star \#10225 - a peculiar object}
\label{sec:10225}

\begin{figure*}[htb]
	\plotone{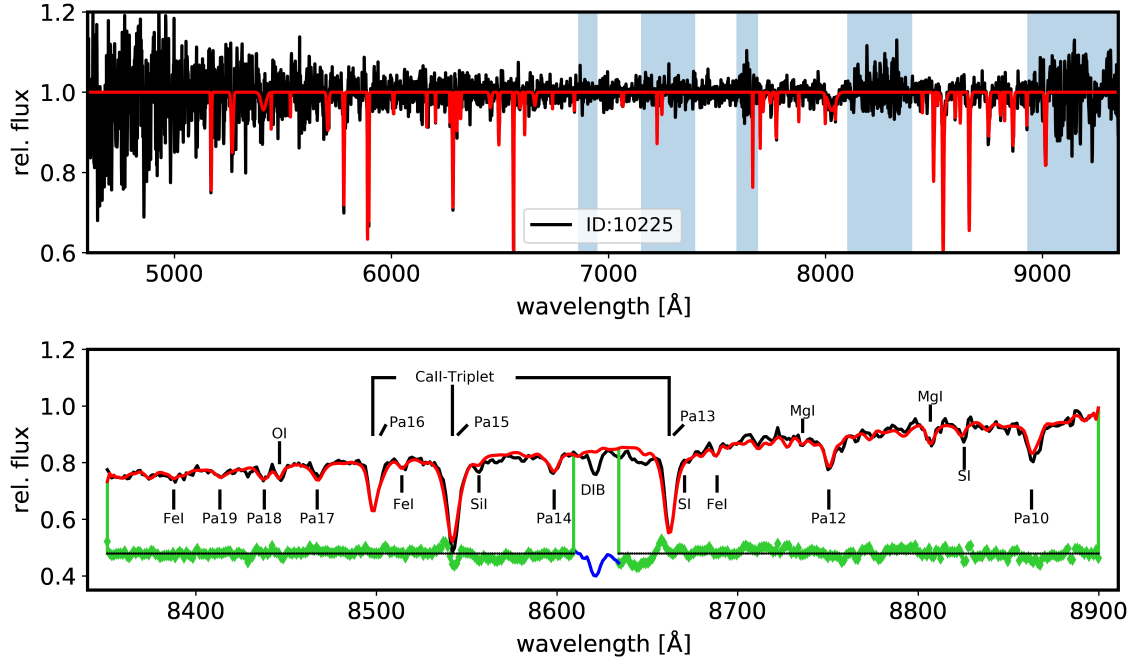}
	\caption{MUSE spectra for star \#10225: The top panel shows the rectified spectrum with the fitted absorption lines in red. The estimated mass based on its location in the CMDs is $4.25~{\rm M}_\odot$. The bottom panel shows the \ion{Ca}{2}-triplet region with the extracted spectrum in black and the best fitting template \citep{Cenarro_01} in red. In green are marked the fit residuals. We marked the important lines used for the spectral typing. The DIB at $8620{\rm \AA}$ is also marked.}
	\label{fig:10225_spec_type}
\end{figure*}

The spectrum of star \#10225 is shown in Fig.~\ref{fig:10225_spec_type}. The weak Balmer lines (H$\beta$ disappears in the noise) suggest a spectral type of F or later. Despite the weak hydrogen lines, the Paschen (Pa)-series is visible up to the P18/19-lines suggesting this star is a giant or supergiant (luminosity class I or II). The lack of nitrogen lines in the \ion{Ca}{2}-triplet region and the appearance of many neutral metals, such as \ion{Fe}{1}, \ion{Si}{1}, and \ion{Mg}{1} (see bottom panel of Fig.~\ref{fig:10225_spec_type}) favors a G-type star. The best fitting template is the one of a G5Ib star, a yellow supergiant. The spectral type of a G5Ib star in combination with the photometric mass of $4.25\,{\rm M}_\odot$ \citep{Zeidler_17} would suggest an age of 10--30~Myr, much older than the age of Wd2 ($\sim 2$\,Myr). Although the location of a star in color-magnitude diagrams (CMDs) defines it as a cluster member, the selection based on photometry alone may be misleading. Wd2 is located in the MW disk and, therefore, foreground interlopers or highly reddened supergiants located behind the cluster are able to contaminate the PMS. \citet{Hur_15} estimated the probability for a field star contamination to be 2.8\%, which is small but not impossible. While the high extinction caused by the \ion{H}{2} region makes the scenario of a background source contaminating the cluster sequence basically impossible, foreground field stars are not excluded.

\begin{figure*}[htb]
	\plotone{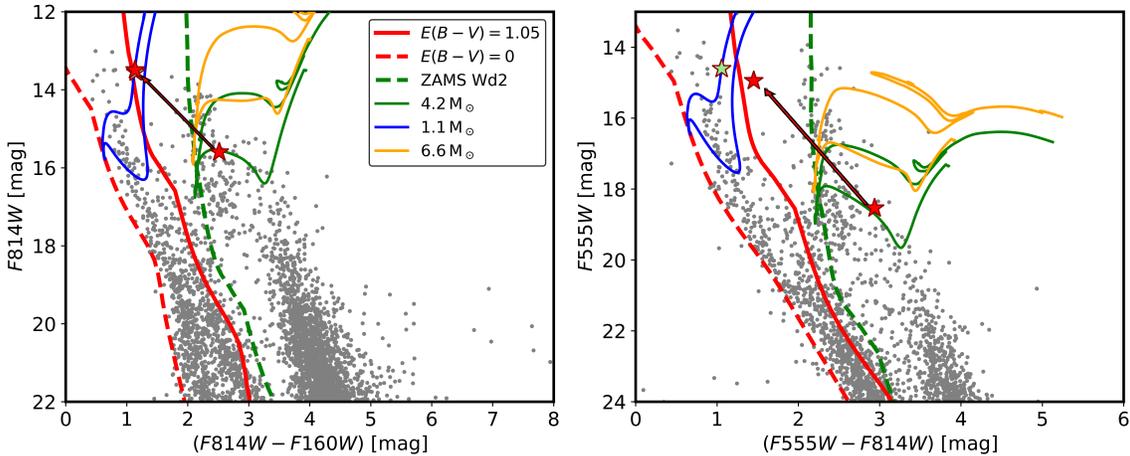}
	\caption{We show the $F814W-F160W$ vs. $F814W$ (left panel) and $F555W-F814W$ vs. $F555W$ (right panel) CMDs. The red (dashed) lines mark the field star sequences as defined in \citet{Hur_15}.The green dashed line is the ZAMS for Wd2 \citep{Zeidler_15}. The green and yellow lines are the evolutionary tracks for the G5 ($4.25\,{\rm M}_\odot$) and B4.5 ($6.6\,{\rm M}_\odot$) cluster member stars, respectively. The red stars mark the loci of the star \#10225 for the star being a cluster member and the star begin a foreground object. The blue line represents the evolutionary track of a $1.1\,{\rm M}_\odot$ star, the best-fitting solution if star \#10225 is a field star. The green star marks the position of  \#10225 if we use an abnormal exinction law of $R_{\rm V} = 2.45$ and  $E(B-V)_{\rm fg} = 1.55\,{\rm mag}$.}
	\label{fig:location_10225}
\end{figure*}

In the following we will discuss the two options: 1) the star is a G5 sub-giant occupying the same region of the CMD as Wd2's PMS, and 2) the star is G5 PMS star:

1) The star is a G5 interloper: despite the low probability \citep[2.8\%,][]{Hur_15}, star \#10225 might be a foreground sub-giant occupying the same locus in the CMD as the Wd2 PMS. \citet{Hur_15} argue that the maximum reddening of the foreground field stars does not exceed $E(B-V)_{\rm fg} = 1.05\,{\rm mag}$ with an $R_{\rm V,fg} = 3.33$ and a distance modulus of 11.8\,mag. As a result, $A_{\rm V,fg} = R_{\rm V,fg} \times E(B-V)_{\rm fg} = 3.50\,{\rm mag}$ is the maximum extinction correction that can be applied to any field star in front of Wd2. In Fig.~\ref{fig:location_10225} we show the $F814W-F160W$ vs. $F814W$ (left panel) and $F555W-F814W$ vs. $F555W$ (right panel) CMDs. The measured and foreground-extinction corrected positions of star \#10225 and the reddening vector are indicated. In the $F814W-F160W$ vs. $F814W$ CMD the best-fitting stellar evolutionary track\footnote{For all isochrones and evolutionary tracks, we used the MESA Isochrones \& Stellar Tracks (MIST)} \citep[blue line,][]{Paxton_11,Paxton_13,Paxton_15,Choi_16,Dotter_16} represents a $1.1\,{\rm M}_\odot$ star and \#10225 would be indeed a G5 sub-giant. In the $F555W-F814W$ vs. $F555W$ the same evolutionary track does not fit the locus of the star. A distance modulus of 14.4\,mag ($d=6.6\,{\rm kpc}$) is necessary to fit the star. We also tested the scenario that there is a dust cloud located in front of this star, which changes the extinction law. To fit the star's position to the evolutionary tracks in both CMDs an abnormal extinction law of $R_{\rm V} = 2.45$, $E(B-V)_{\rm fg} = 1.55\,{\rm mag}$ would be required (green star in Fig.~\ref{fig:location_10225}). Therefore, we can conclude that star \#10225 is not a foreground star.

2) The star is a G5 PMS star: under the assumption that this star is a cluster member the photometric mass is $4.25\,{\rm M}_\odot$ \citep{Zeidler_17}. The stellar evolutionary track for such a stellar mass (see green lines in Fig.~\ref{fig:location_10225}) fits the locus of this star for both CMDs. Late-type PMS stars appear much brighter during their brief period on the turn-on compared to their MS life. This leads to the effect that a low-mass PMS star (green evolutionary track in Fig.~\ref{fig:location_10225}) is similarly bright than a higher-mass early type star (yellow evolutionary track in Fig.~\ref{fig:location_10225}), e.g., star \#10759, which has been classified as a B4.5 star (see Sect.~\ref{sec:Bstars} and Tab.~\ref{tab:spec_stars}). Although their spectral types and masses are different (star \#10759 has a photometric mass of $6.6\,{\rm M}_\odot$) their brightness only differs by 1.27\,mag in $F555W$ and 0.93\,mag in $F814W$. The misclassification of the spectral type as a G5Ib star may have happened due to "veiling". Veiling describes a reduction of the photospheric absorption lines by an accretion continuum in T-Tauri stars. Especially affected are hydrogen lines \citep[e.g., ][ and references therein]{Herczeg_14}, which increases the relative strength of the \ion{Ca}{2}-triplet compared to the Pa-Series. The detection of emission lines as a result of active mass accretion is not possible because of the low S/N.

With the above argumentation in mind, we conclude that star \#10225 is a cluster member with a spectral type of G5 PMS. Although possible, the offset between the evolutionary track and the star's extinction corrected location in the $F555W-F814W$ vs. $F555W$ CMD on the one side and the well fitting locus as a cluster member on the other, makes it unlikely that the star is a foreground interloper.

\section{The radial velocities}
\label{sec:stellar_velocities}
The stellar velocity dispersion of young star clusters is typically on the order of $\sim 5\,{\rm km}\,{\rm s}^{-1}$ \citep[e.g., ][]{Rochau_10,Pang_13,Cottaar_14,Kimiki_18}. To measure these velocities high-resolution spectrographs are typically used (with resolving power of $R \approx 30000$--60000). MUSE was designed to study high-redshift galaxy formation and star-formation in nearby galaxies, whose expected radial velocity (RV) dispersions are on the order of a few $100\,{\rm km}\,{\rm s}^{-1}$. To measure RVs of YMCs in the MW with MUSE we take advantage of the large wavelength range, allowing us to simultaneously measure RVs from a number of strong stellar absorption lines. Together with a statistical approach this allows us to reach the needed accuracy of a few ${\rm km}/{\rm s}$. The procedure is explained in the following section.

\subsection{The stars}
We selected a number of strong stellar absorption lines to measure the RVs: \ion{He}{2}$\lambda4685$,  \ion{Mg}{1}$\lambda\lambda5267,5172,5183$, \ion{He}{2}$\lambda5411$, \ion{He}{1}$\lambda5876$, \ion{He}{2}$\lambda5411$, \ion{He}{1}$\lambda5876$, \ion{Fe}{2}$\lambda6456$, \ion{He}{1}$\lambda6678$, \ion{Ca}{2}$\lambda\lambda8498,4542,8662$. We intentionally did not use the H$\alpha$ and H$\beta$ lines because these lines may be contaminated by outflows and disks, especially for the young cluster stars.

To create a template for each of the lines, we used \texttt{pyspeckit} to fit Voigt-profiles (which is a convolution of a Gaussian and a Lorentzian profile) to the lines. The Voigt profile is defined as follows:

\begin{equation}
V\left(x,\sigma,\gamma\right)=\frac{\Re \left[w\left(z\right)\right]}{\sigma \sqrt{2\pi}},
\label{eq:Voigt}
\end{equation}
where $\sigma$ is the width of the Gaussian and $\gamma$ the width of the Lorentzian. $\Re \left[w\left(z\right)\right]$ is the real part of the Faddeeva function evaluated for:

\begin{equation}
z=\frac{x+\i\gamma}{\sigma \sqrt{2}}.
\label{eq:Faddeeva}
\end{equation}

Similar to what was done in Sect.~\ref{sec:extracted_spectra} we used a $1^{\rm st}$ or $2^{\rm nd}$ order polynomial to model the continuum (depending on each line). We modeled all lines in each of the wavelength regions simultaneously. Using a Voigt profile and a small wavelength range resulted in optimized fits and line parameters. We then used the central wavelength extracted from the NIST library, while for the other parameters (amplitude, $\sigma$, and $\gamma$) we used the results from the above described line fitting. These templates, together with the observed spectra were fed into pPXF, which cross-correlates the template and the measured spectrum. Since our cluster-member stars are young stars possibly containing outflows, we only used the cores of each line. To make sure the determined velocities are not dominated by noise, we adopted a Monte-Carlo (MC) approach, where we repeated the cross correlation 20000 times. For each iteration the uncertainties where added, following a random normal distribution. The mean of the resulting Gaussian distribution was used as the measured RV and the width as its uncertainty. As a first step, we determined the RV of each line individually and applied a sigma clipping in velocity space. Using a $3\sigma$ threshold has proven to be sufficient to remove "outliers" caused by peculiar line shapes (e.g., non-perfect background subtraction, remaining cosmic rays, or a non-perfect wavelength calibration). In a second step, we reprocessed the complete spectrum for a final velocity solution. Again, the mean of the resulting Gaussian distribution was used as the measured RV and the width as RV uncertainty. The radial velocities and their uncertainties are shown in Column 8 and 9 of Tab.~\ref{tab:spec_stars}. The typical RV uncertainty is $\sigma_{\rm RV} = 2.90\,{\rm km}\,{\rm s}^{-1}$.

\begin{figure}[htb]
	\plotone{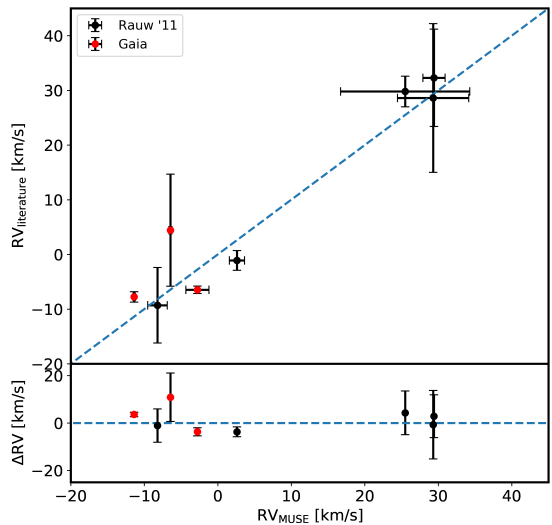}
	\caption{The RV comparison between our RV measurements (x-axis) and the existing literature values (y-axis, \citet{Rauw_11} measurements in black and Gaia measurements in red). The blue line is the theoretical 1:1 relation. The bottom plot shows the difference between our results and the literature values.}
	\label{fig:RV_comparison}
\end{figure}

The limited spectral resolution of MUSE makes it necessary to carefully check whether the resulting RVs are reliable. While looking for binaries \citet{Rauw_11} measured RVs of different stellar lines for a small sample of stars in Wd2. Five of these stars have RV measurements that can be used as a comparison sample. Additionally, for three stars\footnote{Gaia DR2 5255678122073786240:\\ RV (Gaia): $(4.43 \pm 10.24)\,{\rm km}\,{\rm s}^{-1}$; HST catalog ID: 8485 \\ Gaia DR2 5255677920238526336:\\ RV (Gaia): $(-7.77 \pm 0.94)\,{\rm km}\,{\rm s}^{-1}$; HST catalog ID: 15613 \\ Gaia DR2 5255678263835930624: \\ RV (Gaia): $(-6.47 \pm 0.67)\,{\rm km}\,{\rm s}^{-1}$; HST catalog ID: 19296} RV measurements are available in the Gaia data release 2 \citep[DR2, ][]{Gaia_16,Gaia_18}. The overall, error-weighted RV difference is $3.70\,{\rm km}\,{\rm s}^{-1}$.  In Fig.~\ref{fig:RV_comparison} we show the \citet{Rauw_11} measurements (black) and Gaia measurements (red) vs. ours, confirming that reliable RVs with an accuracy of 2--$3\,{\rm km}\,{\rm s}^{-1}$ can be measured with MUSE when multiple stellar lines are used. The typical RV uncertainty of our measurements is $\sigma_{\rm RV} = 2.90\,{\rm km}\,{\rm s}^{-1}$, which is similar to the results found in \citet{Kamann_16}. Their median RV uncertainty is $3.0\,{\rm km}\,{\rm s}^{-1}$ for stars whose spectra have a S/N$>28$, which is comparable to our selected sample. Moreover, in their MUSE study of 500000 spectra of 200000 stars located in 22 globular clusters, \citet{Kamann_18} were able to reach a RV accuracy of $1\,{\rm km}\,{\rm s}^{-1}$ .

\subsection{The gas}
The same procedure can be used to measure the velocities of different gas emission lines. To increase the S/N, for the gas we always combined at least 3 spaxels as extracted from the reduced spectral cubes. We used three different sets of emission lines, the Balmer lines (${\rm H}\alpha$ and ${\rm H}\beta$), the [\ion{N}{2}]$\lambda\lambda6549,6583$ lines, and the [\ion{S}{2}]$\lambda\lambda6717,6731$  lines. We measured the gas velocities in regions not contaminated by bright stars.

\section{Results and discussion}
\label{sec:results}

\subsection{The spectral types -- Cluster members vs. field stars}
To avoid biases, we have so far treated all stars independently of whether they are cluster members or not. To study the internal dynamics of the NB as well as its origin, we need to distinguish between cluster member stars and field stars. Due to the young age of Wd2, we expect the early-type stars to belong to the cluster and not to the field, where the earliest types (O and B) have disappeared, due to their short lifetime \citep[e.g.,][]{Massey_03}. Since the most luminous stars in our data have a spectral ${\rm S/N} \geq 20$, we do not expect to see cluster members with a spectral type much later than an A-type star. The field stars may cover a by far wider dynamic range for their spectral types due to a much lower extinction compared to Wd2 \citep[$A_{\rm V}=6.12\,{\rm mag}$,][]{Zeidler_15}. This high extinction is also the reason why we expect that all field stars detected in our MUSE spectra are located in front of Wd2.

\begin{figure*}[htb]
	\plotone{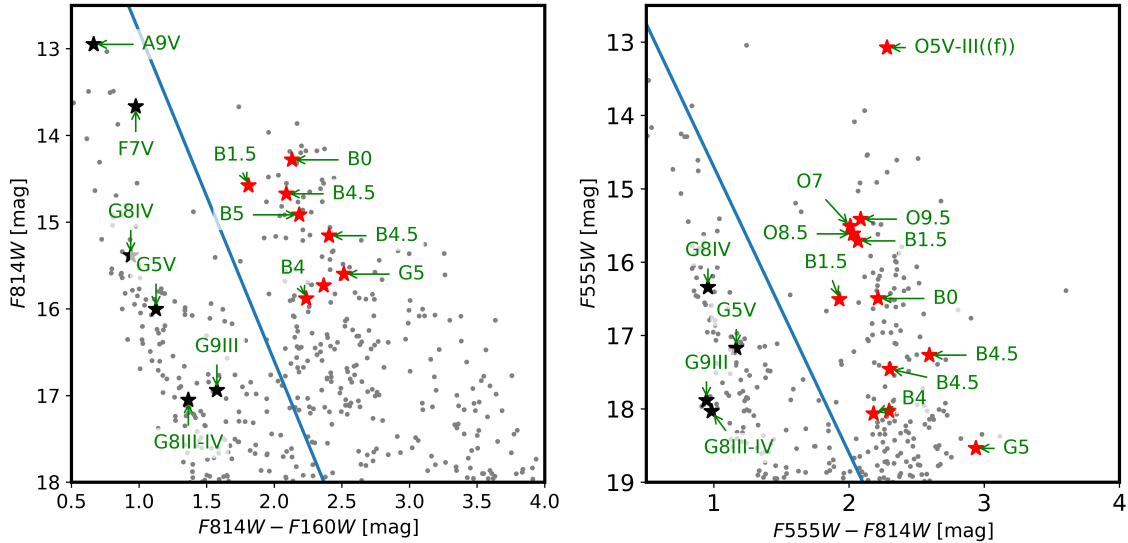}
	\caption{The  $F814W-F160W$ vs. $F814W$ (left panel) and $F555W-F814W$ vs. $F555W$ (right panel) CMDs of all stars detected in our HST catalog (gray small dots). The blue line shows the photometrically estimated separation between the cluster members and field stars \citep{Zeidler_15}. The stars in the NB whose spectral type was derived from our MUSE data are marked with asterisks.}
	\label{fig:CMD+spec}
\end{figure*}

We used our HST photometric catalog to create multiple CMDs (see Fig.~\ref{fig:CMD+spec}) and marked all stars located in the NB, of which we determined their spectral type (red and black asterisks in Fig.~\ref{fig:CMD+spec}). In \citet{Zeidler_15} we showed that the field stars and the cluster members can be well separated by a linear cut in color-magnitude space, represented by the blue line in Fig.~\ref{fig:CMD+spec}. Comparing spectral types and locations in the CMDs, we see that the earliest spectral type among field stars is A9V, while OB stars appear to be cluster members except star \#10255 (see Sect.~\ref{sec:10225} for a detailed discussion).

This analysis is based only on stars in the NB. It shows that, for a given magnitude range, the spectral types may be used as an additional indicator to separate cluster members from field stars, which is important at loci where the two populations (cluster and field) are not very well separated.

\subsection{The radial velocity distribution of Wd2}
 
\subsubsection{The stars}

For the NB, there are 24 stars suited to measure RVs. To increase the sample for further statistical analyses of the stellar velocity distribution, we added 48 stars from the remaining cluster area including stars located in some of the recently obtained long exposures. These stars were selected based on their good-quality extracted spectra. An overview over all stars used for this work is presented in Tab.~\ref{tab:spec_stars}.

\begin{figure*}[htb]
	\plotone{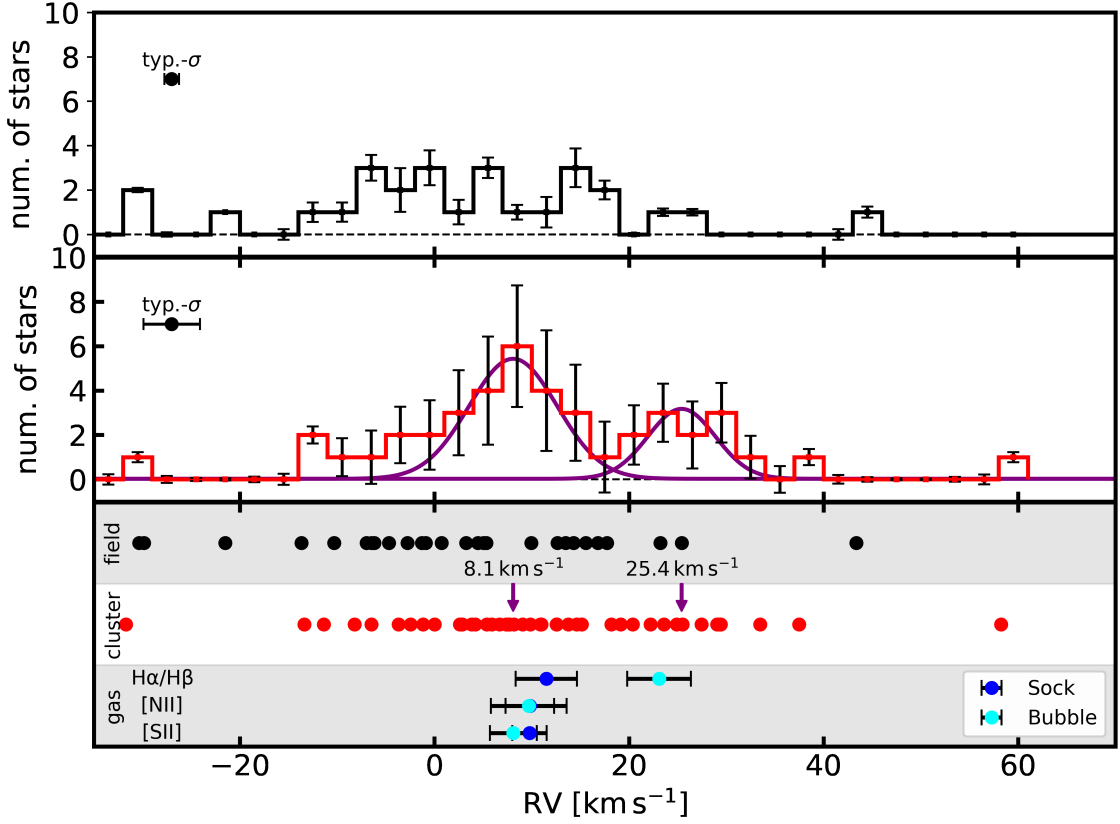}
	\caption{\textbf{Top panel:} The RV distribution of the field stars (black). We show the typical velocity uncertainty for a single star, obtained from the field stars only ($\sigma = 0.76\,{\rm km}/{\rm s}$). \textbf{Middle panel:} The RV distribution of the cluster stars (red). We show the typical velocity uncertainty for a single star ($\sigma = 2.90\,{\rm km}/{\rm s}$), as well as the number uncertainty per RV bin. The purple curves show the most probable result of the MCMC simulation for the bimodal distribution. \textbf{Bottom panel:} The locus of the cluster stars (red) and the field stars (black) in RV space. The mean RVs of $(8.10 \pm 1.53)\,{\rm km}\,{\rm s}^{-1}$ and $(25.41 \pm 1.57)\,{\rm km}\,{\rm s}^{-1}$ of the bimodal distribution are indicated by the purple arrows. We also show the median gas velocities measured from the Balmer lines ($(11.50 \pm 3.15)\,{\rm km}\,{\rm s}^{-1}$ and $(23.09 \pm 3.27)\,{\rm km}\,{\rm s}^{-1}$), the [\ion{N}{2}]$\lambda\lambda6549,6583$ lines ($(9.81 \pm 2.50)\,{\rm km}\,{\rm s}^{-1}$ and $(9.70 \pm 3.90)\,{\rm km}\,{\rm s}^{-1}$), and the [\ion{S}{2}]$\lambda\lambda6717,6731$ lines ($(9.77 \pm 1.76)\,{\rm km}\,{\rm s}^{-1}$ and $(8.22 \pm 2.42\,{\rm km}\,{\rm s}^{-1}$). The velocities are always given for "the Sock" and the NB, respectively.}
	\label{fig:RV}
\end{figure*}

We analyzed the RV distribution throughout the Wd2 area including all 72 stars (see Tab.~\ref{tab:spec_stars}, faintest star: $V = 20.53\,{\rm mag}$), 44 of which are cluster members, based on their location in the CMDs of Fig.~\ref{fig:CMD+spec}. In the lower panel of Fig.~\ref{fig:RV} we show the RV distribution of the cluster members (red) and the field stars (black), where the binsize is set equal to the typical velocity uncertainty of $\sim 3\,{\rm km}/{\rm s}$. We propagated the individual measured RV uncertainty per star to a cumulative uncertainty on the number of stars per velocity bin, assuming they are Bernoulli distributed. As a result, also bins with no stars may have a non-zero uncertainty.

To quantify the distribution and to test whether it is statistically significant, we applied a Markov Chain Monte Carlo (MCMC) method on a combination of two Gaussian distributions and a common offset including the number uncertainties per RV bin. We ran $8\times 10^5$ random draws and confirmed the bimodal distribution. The most probable RV of each of the two peaks is $(8.10 \pm 1.53)\,{\rm km}\,{\rm s}^{-1}$ and $(25.41 \pm 1.57)\,{\rm km}\,{\rm s}^{-1}$. The $1\sigma$ widths of the two peaks are $(4.52 \pm 1.78)\,{\rm km}\,{\rm s}^{-1}$ and $(3.46 \pm 1.29)\,{\rm km}\,{\rm s}^{-1}$. To exclude a bias for the choice of the priors, we also ran the MCMC fit using a single Gaussian and a combination of three Gaussians, both with a common offset. Each MCMC run did not converge to a reliable solution.

We tested if the cluster and field populations are following different distributions. First, we tried to fit the same distribution as for the cluster members to the field population. This MCMC run did not converge to a reliable solution. Second, we ran a Kolmogorov-Smirnov (K.S.) test with the null hypothesis that the two populations are from the same parental sample. Based on the obtained p-value of 0.04, we can reject the null hypothesis. Both tests confirm that the field and the cluster population are indeed different.
	
\begin{figure*}[htb]
	\plotone{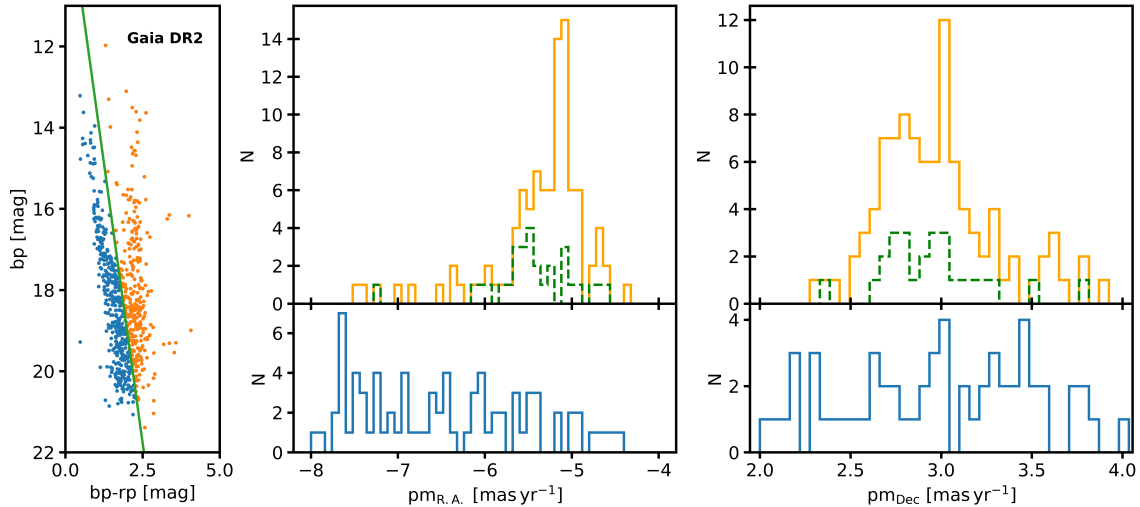}
	\caption{\textbf{Left panel:} The Gaia CMD toward Wd2. The green line separates likely cluster members (orange) from the likely field stars (blue). In total 421 field stars and 302 cluster members are detected. \textbf{Center panel:} The proper motion distribution in right ascension of all stars that have an uncertainty less than $0.2\,{\rm mas\,yr}^{-1}$. The top panel shows the cluster members while the bottom panel shows the field stars.  \textbf{Right panel:} The proper motion distribution in declination of all stars that have an uncertainty less than $0.2\,{\rm mas\,yr}^{-1}$ showing a bimodal distribution. The top panel shows the cluster members while the bottom panel shows the field stars. The dashed green histograms show the distribution of the stars that also have MUSE RVs.}
	\label{fig:pm}
\end{figure*}

To avoid being biased by the member selection performed with our HST catalog we used the Gaia DR2 as a third, independent test. Due to the high extinction and crowding toward Wd2 the parallaxes and proper motions of the majority of the stars in that region still have large uncertainties. We divided the stars in likely foreground field stars and likely cluster members based on the Gaia DR2 photometry (see left panel of Fig.~\ref{fig:pm}). We then selected all probable cluster member stars, for which the proper motion uncertainty is less than $0.2\,{\rm mas\,yr}^{-1}$ (corresponding to $3.9\,{\rm km\,s}^{-1}$ in declination and $2.1\,{\rm km\,s}^{-1}$ in right ascension at a distance of 4.16\,kpc, selecting 111 and 109 stars, respectively). We show the proper motions in right ascension and declination in the center and right panel of Fig.~\ref{fig:pm}, respectively. The proper motion distribution in declination shows the same double peak as the RV distribution (see Fig.~\ref{fig:RV}), supporting the bimodality of Wd2's stellar population. In right ascension a double peak is not apparent. The reason for this may be a very small difference between the two peaks, not detectable by the current accuracy of the Gaia data. Another possibility is that  due to the North - South orientation of the two clumps \citep{Hur_15,Zeidler_15}, such a bimodal distribution does not exist in this direction. Gaia DR3 will shade light on this. We cross matched the cluster stars for which we have RV measurements with the one for which we have proper motions. Their distribution is shown as green dashed histogram in Fig.~\ref{fig:pm}. Both histograms show a bimodal distribution. The number of stars is not high enough to obtain a significant result from cross matching the two peaks individually. As in RV space, the velocity distribution of the field stars (bottom panels of Fig.~\ref{fig:pm}) shows a uniform distribution with a similar velocity range as the cluster members. This excludes the possibility of using the proper motions for an additional criterion for the membership selection as it was done, e.g., the Orion Star Forming Complex \citep{Kounkel_2018}.

Comparing the velocity distribution of the RVs and in declination, in addition to the K.S. test, the MCMC fit, and the cross-matched histograms, we conclude that the Wd2 stars show a bimodal velocity distribution.

\begin{figure*}[htb]
	\plottwo{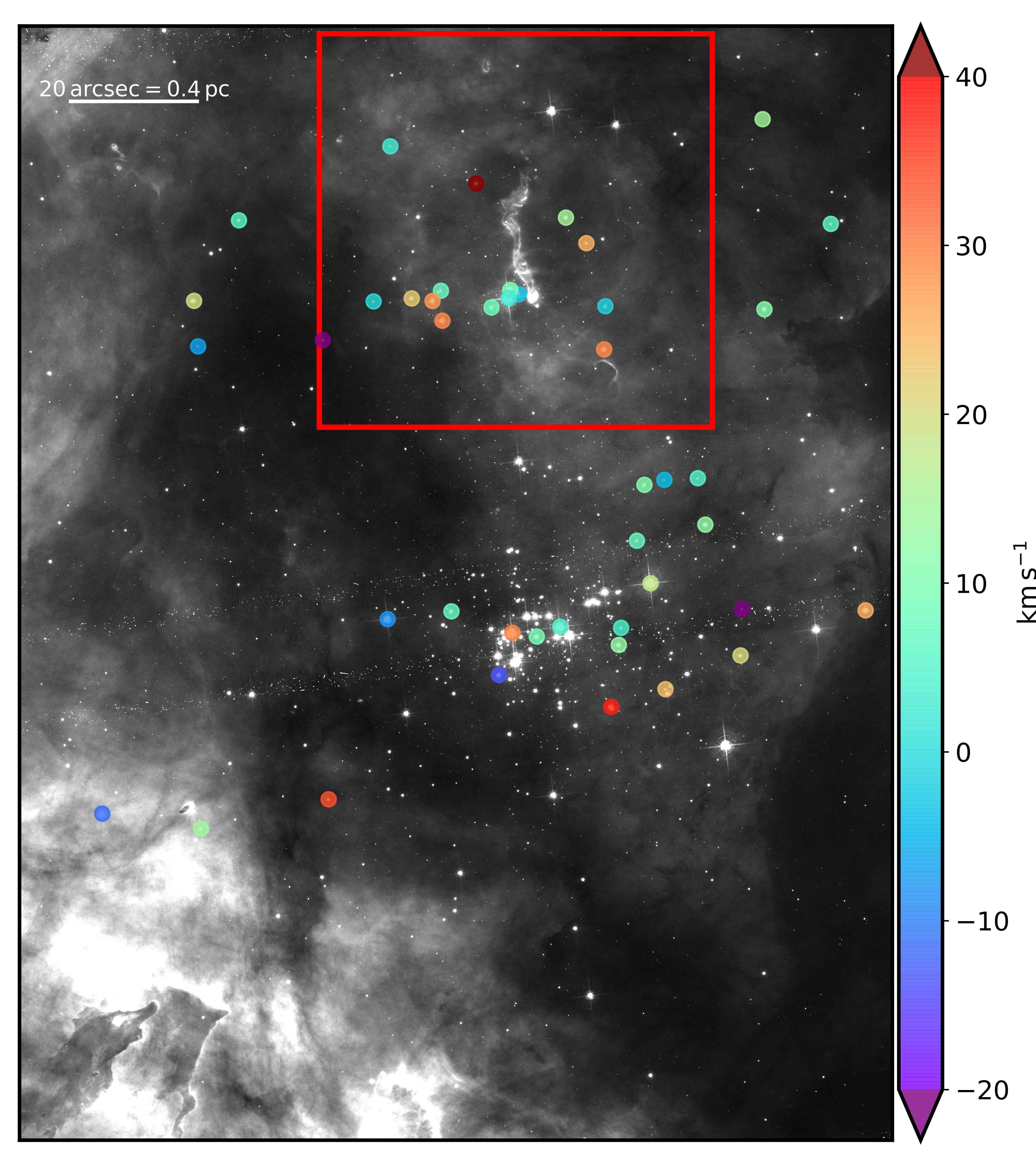}{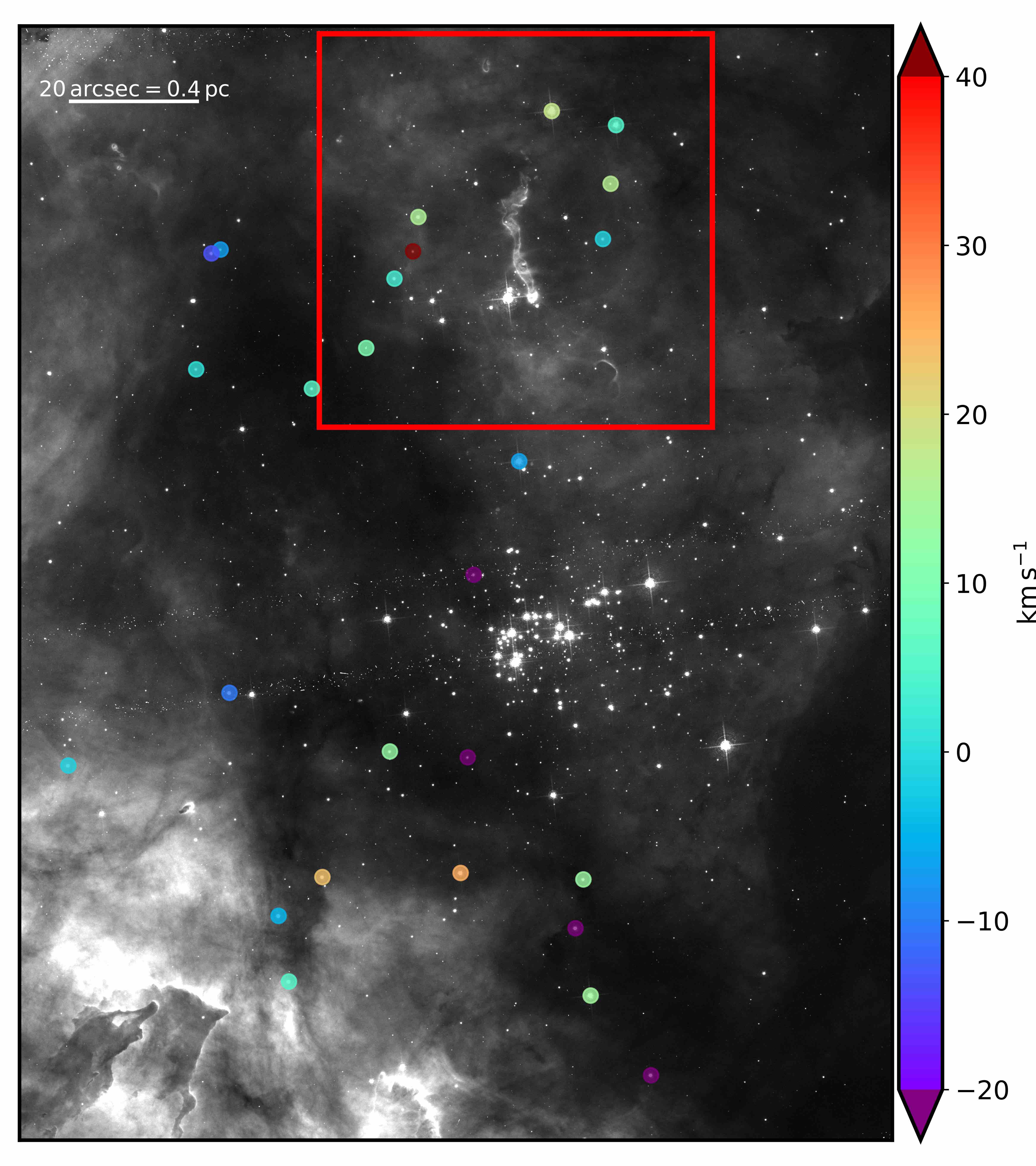}
	\caption{\textbf{Left:} The cluster-member stars in the NB for which we measured the RVs. The colorbar shows the measured RVs. To see more details we used the HST $F658N$ (${\rm H}\alpha$) image as background. \textbf{Right:}  The field stars for which we measured the RVs.}
\label{fig:vel_dist_stars}
\end{figure*}

In Fig.~\ref{fig:vel_dist_stars} we show the location of all cluster members (left panel) and all field stars (right panel) color-coded by their respective RVs. The red square marks the $64''\times64''$ region around the NB similar to Fig.~\ref{fig:mosaic}. The selection of stars is fairly uniform. There is no hint of a spatial correlation, especially in the field stars. This shows that there are no calibration artifacts left, which may influence the accuracy of our results.

Based on CO($J=1$--2) NANTEN2 sub-millimeter observations of RCW49, \citet{Furukawa_09} argued for the formation of Wd2 having been triggered by a collision between two molecular clouds ($\sim4\,$Myr ago) as a viable scenario. The CO velocity profile of the whole region revealed that there are two gas clouds moving with an RV of $0^{+9}_{-11}\,{\rm km}\,{\rm s}^{-1}$, located in front of Wd2, and $16^{+5}_{-4}\,{\rm km}\,{\rm s}^{-1}$, respectively, located in the background of Wd2. The velocity difference of the two stellar components ($\Delta v_\star = (17.30\pm2.17)\,{\rm km}\,{\rm s}^{-1}$) and the velocity difference of the two gas clouds ($\Delta v_{\rm gas} = 16^{+10.3}_{-11.7}\,{\rm km}\,{\rm s}^{-1}$) are similar. Although the mean velocities of both clouds are lower than the velocities of the two stellar components they might be connected. A detailed RV analysis of all stars and the gas is necessary to establish a connection between the stellar RV distribution and a probable formation scenario of Wd2. We will address this in a future work analyzing the complete dataset.

In a recent study, \citet{Kimiki_18} measured a velocity dispersion of $\le 9.1\,{\rm km}\,{\rm s}^{-1}$ or 41 well-constrained O-type stars in the Trumpler~14 cluster in the Carina Nebula \citep[e.g., ][and references therein]{Smith_08b}. \citet{Rochau_10} used HST/WFPC2 \citep{WFPC2} observations to measure proper motions of NGC~3603 and determined a 1D velocity dispersion of $(4.5 \pm 0.8)\,{\rm km}\,{\rm s}^{-1}$. \citet{Pang_13} repeated the analysis with the same dataset but for each tangential component individually resulting in an slightly increased velocity dispersion ($\sigma_{x} = (4.8 \pm 0.5)\,{\rm km}\,{\rm s}^{-1}$ and $\sigma_{y} = (6.5 \pm 0.5)\,{\rm km}\,{\rm s}^{-1}$), which was most likely caused by a different selection method and detection uncertainty correction. Comparing these studies to the $1\sigma$ widths of the two RV distributions in Wd2 ($(4.52 \pm 1.78)\,{\rm km}\,{\rm s}^{-1}$ and $(3.46 \pm 1.29)\,{\rm km}\,{\rm s}^{-1}$) we can conclude that the RV dispersions derived for the two components of Wd2 are comparable with measurements for other Galactic YMCs. We will perform a detailed velocity dispersion analysis and estimates of the dynamical cluster mass in a future paper. 

\subsubsection{The gas}
\begin{figure*}
	\plotone{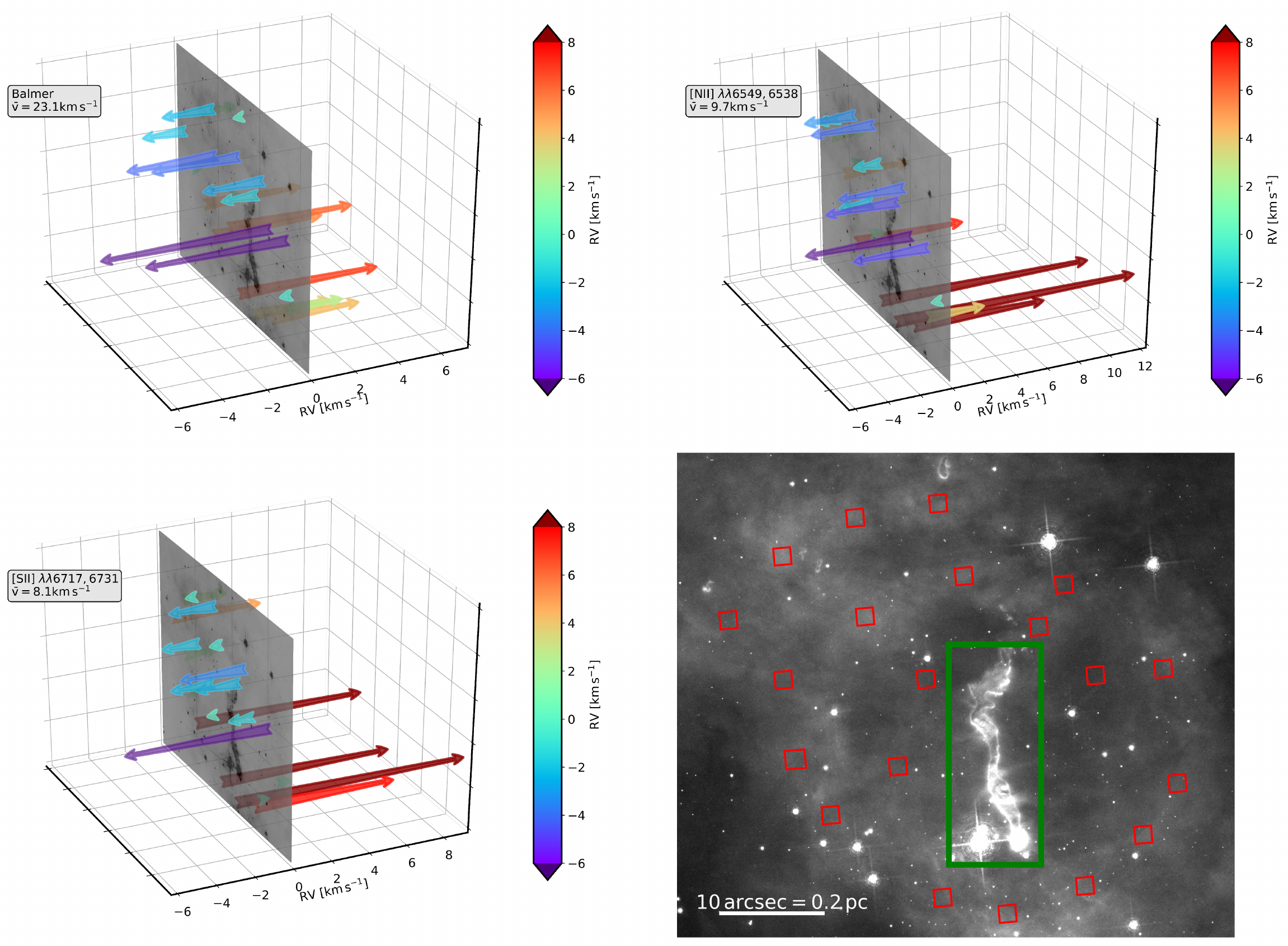}
	\caption{The gas velocities of the NB derived from the Balmer lines, the [\ion{N}{2}]$\lambda\lambda6549,6583$ lines, and the [\ion{S}{2}]$\lambda\lambda6717,6731$ lines. We use the respective median velocity (indicated in the top left corner) as reference frame. The arrows represent the line-of-sight velocities. We chose the HST $F658N$ (${\rm H}\alpha$) filter as reference image. The typical velocity uncertainty is $0.7\,{\rm km}\,{\rm s}^{-1}$, derived in the same manner as the stellar RVs and their uncertainties. In the bottom right panel we show the H$\alpha$ image face-on and marked the 20 fields where we measured the RVs with red squares. The green area marks "the Sock" (see Fig.~\ref{fig:vel_gas_sock}). North is up, East is to the left.}
	\label{fig:vel_gas_NB}
\end{figure*}

\begin{figure*}
	\plotone{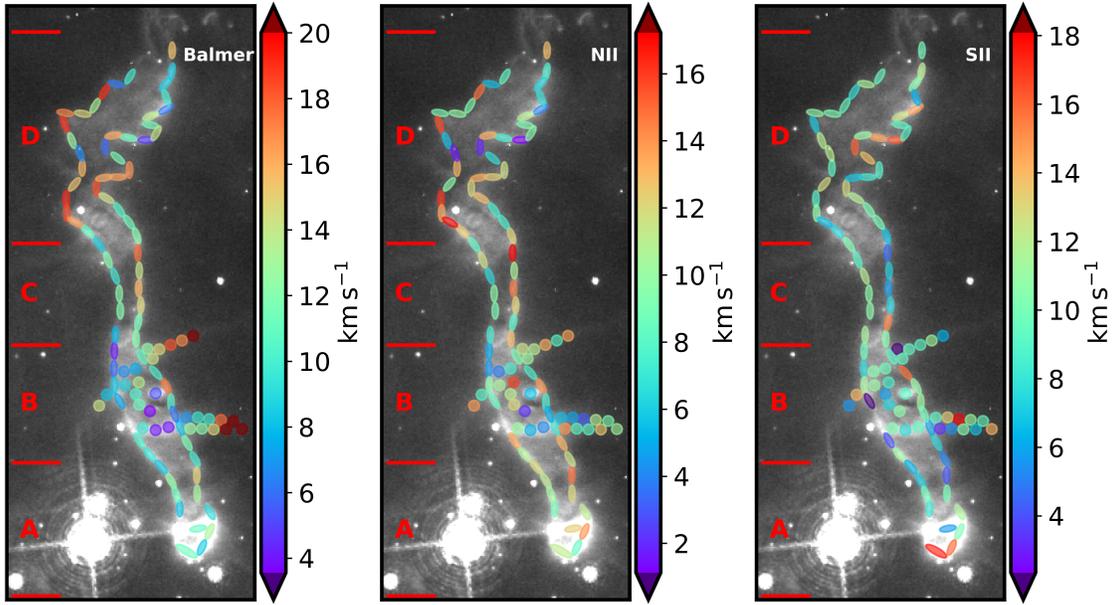}
	\caption{The gas velocities of "the Sock" (marked in green in Fig.~\ref{fig:vel_gas_NB}) derived from the Balmer lines, the [\ion{N}{2}]$\lambda\lambda6549,6583$ lines, and the [\ion{S}{2}]$\lambda\lambda6717,6731$ lines. We use the respective median velocity (indicated in the top left corner) as reference frame. The four main regions are marked on the left. To see more details we selected the HST $F658N$ (${\rm H}\alpha$) image as background. The typical velocity uncertainty is $0.4\,{\rm km}\,{\rm s}^{-1}$, derived in the same manner than the stellar RVs. The luminous star to the South-East has a spectral type of O5III-V((f)) \citep{Vargas_Alvarez_13}. Each panel shows $9" \times 22"$. North is up, East is to the left.}
	\label{fig:vel_gas_sock}
\end{figure*}

To study the velocity distribution of the gas we chose 20 fields across the NB with a typical size of $1.715 \times 1.715\,{\rm arcsec}^2 $ or $8.575 \times 8.575\,{\rm px}^2$ (see Fig.~\ref{fig:vel_gas_NB}). For "the Sock" we used 102 smaller, elliptically-shaped fields to cover the rims on both sides, as well as visible features of the gas (e.g., a possible bow shock). The typical size of an ellipse is $0.12\,{\rm arcsec}^{2}$, which corresponds to $\sim 3\,$px$^2$. All of the covered areas where chosen in a way that excludes stars. In Fig.~\ref{fig:vel_gas_NB} and Fig.~\ref{fig:vel_gas_sock} we present the results for the distributions in gas velocity for the NB and the rims of "the Sock", respectively.

The median velocity of each element of the NB and "the Sock" is reported in the bottom panel of Fig.~\ref{fig:RV}. The median gas velocities are, within the uncertainties in agreement with the stellar velocities of the blue shifted peak, thus evidencing a spatial connection between gas and stars, with the latter being formed in the cavity of the gas cloud. The median gas velocity of the NB as derived from the Balmer lines appears to be $\sim 11\,{\rm km}\,{\rm s}^{-1}$ higher than all the other RV measurements, indicating a more complicated velocity structure. The whole NB appears to be slightly red-shifted, indicating that it is moving away from us.

To better compare the velocities of the various components, we subtracted the median velocity for each of the three elements in Fig.~\ref{fig:vel_gas_NB} and Fig.~\ref{fig:vel_gas_sock}. The median velocity of the gas is also presented in the bottom panel of Fig.~\ref{fig:RV}. The RVs of the gas in the NB indicate rotation probably caused by the winds from the massive O-stars inside the NB. The South-East side moves away from us while the North-West side moves toward us (see Fig.~\ref{fig:vel_gas_NB}).  We divided "the Sock" in four major regions labeled A to D as we move from South to North (see Fig.~\ref{fig:vel_gas_sock}). Looking at the gas RVs of "the Sock", derived by analyzing the Balmer lines, it shows a "twisted" velocity profile:

\begin{itemize}
	\item [A)] The stars located East of the tip show similar velocities as "the Sock", which is a strong indication that they are spatially connected. We suggest that the slightly arch-shaped morphology of "the Sock" is caused by the O5III-V((f)) star. Due to insufficient S/N we were unable to extract useful spectra for the two sources directly located at the tip. The left and the right rims have similar velocities.
	\item [B)]  There appears to be a bow-shock pointing South-East with a star embedded in the gas or located behind "the Sock" and therefore not visible. The hydrogen lines of the tail of the bow-shock are more red-shifted than the tip of the bow-shock, which cannot be detected in [\ion{S}{2}] and [\ion{N}{2}].
	\item[C)] The right rim appears to be red-shifted compared to the left side which suggests rotation or expansion. This profile is reversed when looking at the [\ion{S}{2}] lines
	\item[D)] The northern end shows a chaotic structure both in velocity space and in real space, which might be caused by some external disturbance or by the interaction of "the Sock" with the edge on the NB.
\end{itemize}

Overall, the RV profile of "the Sock", derived using the Balmer and [\ion{N}{2}] lines, appears to be twisted with a chaotic northern end. The RV distribution appears more uniform using the [\ion{S}{2}] lines, indicating a complex velocity profile that needs to be further investigated. Deriving gas abundance and temperatures including the influence of all the surrounding stellar sources not only of the NB but of the whole Wd2 region is left for a future paper once the full dataset including the long exposures is fully reduced and analyzed.

\section{Summary and conclusions}
\label{sec:conclusions}

In this work we have presented the first results of the combined study of VLT/MUSE integral field spectroscopy and high-resolution HST photometry of the young massive cluster Wd2 with the aim of studying its stellar and gaseous components in great detail. We focused on a field North of the main cluster, which we call the Northern Bubble, a cavity blown into the remaining gas of the parental gas cloud. In its center, a pillar-like structure or jet-like object, "the Sock", is present (see Fig.~\ref{fig:NB}).

We extracted spectra for 17 stars located in the NB, which are suitable for a spectral classification, using the python based code \pampelmuse~\citep{Kamann_13}. Depending on the spectral type, we cross-matched libraries of the \ion{Ca}{2}-triplet region with the observed spectra (stellar type: A9 and later) or used EW ratios of helium and hydrogen lines (O and B stars). These methods give comparable results to the more traditional way of by-eye comparison with template spectra. With this method the majority of stars can be automatically classified, which will become important as soon as we have the complete dataset of $\gtrapprox 1000$ spectra.

We added another 2 O and 5 B-type stars to the 37 already known OB-type stars \citep{Moffat_91,Vargas_Alvarez_13} in Wd2.

Using strong stellar absorption lines, we derived stellar RVs of 72 stars throughout the Wd2 cluster area with an accuracy of $2.9\,{\rm km\,s}^{-1}$ . The cluster member stars follow a bimodal velocity distribution centered on  $(8.10 \pm 1.53)\,{\rm km}\,{\rm s}^{-1}$ and $(25.41 \pm 1.57)\,{\rm km}\,{\rm s}^{-1}$ (see Fig.~\ref{fig:RV}) with a dispersion of $(4.52 \pm 1.78)\,{\rm km}\,{\rm s}^{-1}$ and $(3.46 \pm 1.29)\,{\rm km}\,{\rm s}^{-1}$, respectively. The dispersions are comparable to those of other Galactic YMCs, such as NGC~3603 \citep{Rochau_10,Pang_13} or Trumpler~14 in the Carina Nebula \citep{Kimiki_18}. The bimodal distribution is also seen in the proper motions of the  Gaia DR2 photometric catalog.

These first results show that it is indeed possible to extract stellar spectra in YMCs from MUSE data to an accuracy where it is possible to estimate the velocity dispersion of PMS stars. We also showed that we can analyze the gas and stars and confirm whether the different components are actually spatially connected.

\acknowledgments
We thank Erik Tollerud for fruitful discussions, suggestions, and ideas about spectral fitting methods and procedures. We thank Adam Ginsburg for helpful input. We thank the anonymous referee for their helpful comments to improve this paper.

This work has made use of data from the European Space Agency (ESA) mission {\it Gaia} (\url{https://www.cosmos.esa.int/gaia}), processed by the {\it Gaia} Data Processing and Analysis Consortium (DPAC, \url{https://www.cosmos.esa.int/web/gaia/dpac/consortium}). Funding for the DPAC has been provided by national institutions, in particular the institutions participating in the {\it Gaia} Multilateral Agreement.

E.K.G., and A.P. acknowledge support by Sonderforschungsbereich 881 (SFB 881, "The Milky Way System") of the German Research Foundation, particularly via subproject B5.

MC acknowledges the INFN (Iniziativa specifica TAsP).

These observations are associated with program \#14807. Support for program \#14807 was provided by NASA through a grant from the Space Telescope Science Institute. This work is based on observations obtained with the NASA/ESA \textit{Hubble} Space Telescope, at the Space Telescope Science Institute, which is operated by the Association of Universities for Research in Astronomy, Inc., under NASA contract NAS 5-26555.

\software{PampleMuse \citep{Kamann_13}, ESORex \citep{Freudling_13}, pyspeckit \citep{Ginsburg_11}, MUSE pipeline \citep[v.2.0.1][]{Weilbacher_12,Weilbacher_14}, Astropy \citep{Astropy_18}, Matplotlib \citep{Hunter_07}, pPXF \citep{Cappellari_04,Cappellari_17}}

\facilities{VLT:Yepun (MUSE), HST(WFC3,ACS), Gaia}

\bibliographystyle{aasjournal}
\bibliography{Wd2_bibliography}

\begin{thebibliography}{}
\expandafter\ifx\csname natexlab\endcsname\relax\def\natexlab#1{#1}\fi
\providecommand{\url}[1]{\href{#1}{#1}}

\bibitem[{{Avila}(2017)}]{ACS}
{Avila}, R.~J. 2017, {Advanced Camera for Surveys Instrument Handbook for Cycle
  25 v. 16.0}

\bibitem[{{Bacon} {et~al.}(2010){Bacon}, {Accardo}, {Adjali}, {Anwand},
  {Bauer}, {Biswas}, {Blaizot}, {Boudon}, {Brau-Nogue}, {Brinchmann},
  {Caillier}, {Capoani}, {Carollo}, {Contini}, {Couderc}, {Daguis{\'e}},
  {Deiries}, {Delabre}, {Dreizler}, {Dubois}, {Dupieux}, {Dupuy}, {Emsellem},
  {Fechner}, {Fleischmann}, {Fran{\c c}ois}, {Gallou}, {Gharsa}, {Glindemann},
  {Gojak}, {Guiderdoni}, {Hansali}, {Hahn}, {Jarno}, {Kelz}, {Koehler},
  {Kosmalski}, {Laurent}, {Le Floch}, {Lilly}, {Lizon}, {Loupias}, {Manescau},
  {Monstein}, {Nicklas}, {Olaya}, {Pares}, {Pasquini}, {P{\'e}contal-Rousset},
  {Pell{\'o}}, {Petit}, {Popow}, {Reiss}, {Remillieux}, {Renault}, {Roth},
  {Rupprecht}, {Serre}, {Schaye}, {Soucail}, {Steinmetz}, {Streicher}, {Stuik},
  {Valentin}, {Vernet}, {Weilbacher}, {Wisotzki}, \& {Yerle}}]{Bacon_10}
{Bacon}, R., {Accardo}, M., {Adjali}, L., {et~al.} 2010, in \procspie, Vol.
  7735, Ground-based and Airborne Instrumentation for Astronomy III, 773508

\bibitem[{{Banerjee} \& {Kroupa}(2015)}]{Banerjee_15a}
{Banerjee}, S., \& {Kroupa}, P. 2015, \mnras, 447, 728

\bibitem[{{Cappellari}(2017)}]{Cappellari_17}
{Cappellari}, M. 2017, \mnras, 466, 798

\bibitem[{{Cappellari} \& {Emsellem}(2004)}]{Cappellari_04}
{Cappellari}, M., \& {Emsellem}, E. 2004, \pasp, 116, 138

\bibitem[{{Cenarro} {et~al.}(2001){Cenarro}, {Cardiel}, {Gorgas}, {Peletier},
  {Vazdekis}, \& {Prada}}]{Cenarro_01}
{Cenarro}, A.~J., {Cardiel}, N., {Gorgas}, J., {et~al.} 2001, \mnras, 326, 959

\bibitem[{{Choi} {et~al.}(2016){Choi}, {Dotter}, {Conroy}, {Cantiello},
  {Paxton}, \& {Johnson}}]{Choi_16}
{Choi}, J., {Dotter}, A., {Conroy}, C., {et~al.} 2016, \apj, 823, 102

\bibitem[{{Cignoni} {et~al.}(2009){Cignoni}, {Sabbi}, {Nota}, {Tosi},
  {Degl'Innocenti}, {Moroni}, {Angeretti}, {Carlson}, {Gallagher}, {Meixner},
  {Sirianni}, \& {Smith}}]{Cignoni_09}
{Cignoni}, M., {Sabbi}, E., {Nota}, A., {et~al.} 2009, \aj, 137, 3668

\bibitem[{{Cottaar} \& {H{\'e}nault-Brunet}(2014)}]{Cottaar_14}
{Cottaar}, M., \& {H{\'e}nault-Brunet}, V. 2014, \aap, 562, A20

\bibitem[{{Dotter}(2016)}]{Dotter_16}
{Dotter}, A. 2016, \apjs, 222, 8

\bibitem[{{Dressel}(2018)}]{WFC3}
{Dressel}, L. 2018, {Wide Field Camera 3 Instrument Handbook v. 10.0}
  (Baltimore: STScI)

\bibitem[{{Freudling} {et~al.}(2013){Freudling}, {Romaniello}, {Bramich},
  {Ballester}, {Forchi}, {Garc{\'{\i}}a-Dabl{\'o}}, {Moehler}, \&
  {Neeser}}]{Freudling_13}
{Freudling}, W., {Romaniello}, M., {Bramich}, D.~M., {et~al.} 2013, \aap, 559,
  A96

\bibitem[{{Fukui} {et~al.}(2014){Fukui}, {Ohama}, {Hanaoka}, {Furukawa},
  {Torii}, {Dawson}, {Mizuno}, {Hasegawa}, {Fukuda}, {Soga}, {Moribe},
  {Kuroda}, {Hayakawa}, {Kawamura}, {Kuwahara}, {Yamamoto}, {Okuda}, {Onishi},
  {Maezawa}, \& {Mizuno}}]{Fukui_14}
{Fukui}, Y., {Ohama}, A., {Hanaoka}, N., {et~al.} 2014, \apj, 780, 36

\bibitem[{{Furukawa} {et~al.}(2009){Furukawa}, {Dawson}, {Ohama}, {Kawamura},
  {Mizuno}, {Onishi}, \& {Fukui}}]{Furukawa_09}
{Furukawa}, N., {Dawson}, J.~R., {Ohama}, A., {et~al.} 2009, \apjl, 696, L115

\bibitem[{{Gaia Collaboration} {et~al.}(2018){Gaia Collaboration}, {Brown},
  {Vallenari}, {Prusti}, {de Bruijne}, {Babusiaux}, \&
  {Bailer-Jones}}]{Gaia_18}
{Gaia Collaboration}, {Brown}, A.~G.~A., {Vallenari}, A., {et~al.} 2018, ArXiv
  e-prints, arXiv:1804.09365

\bibitem[{{Gaia Collaboration} {et~al.}(2016){Gaia Collaboration}, {Prusti},
  {de Bruijne}, {Brown}, {Vallenari}, {Babusiaux}, {Bailer-Jones}, {Bastian},
  {Biermann}, {Evans}, \& et~al.}]{Gaia_16}
{Gaia Collaboration}, {Prusti}, T., {de Bruijne}, J.~H.~J., {et~al.} 2016,
  \aap, 595, A1

\bibitem[{{Gennaro} {et~al.}(2011){Gennaro}, {Brandner}, {Stolte}, \&
  {Henning}}]{Gennaro_11}
{Gennaro}, M., {Brandner}, W., {Stolte}, A., \& {Henning}, T. 2011, \mnras,
  412, 2469

\bibitem[{{Ginsburg} \& {Mirocha}(2011)}]{Ginsburg_11}
{Ginsburg}, A., \& {Mirocha}, J. 2011, {PySpecKit: Python Spectroscopic
  Toolkit}, Astrophysics Source Code Library, , , ascl:1109.001

\bibitem[{{Gonzaga} \& {Biretta}(2010)}]{WFPC2}
{Gonzaga}, S., \& {Biretta}, J. 2010, {in HST WFPC2 Data Handbook, v. 5.0, ed.}
  (Baltimore: STScI)

\bibitem[{{Gray} \& {Corbally}(2009)}]{Gray_09}
{Gray}, R.~O., \& {Corbally}, J., C. 2009, {Stellar Spectral Classification}
  (Princeton University Press)

\bibitem[{{Herczeg} \& {Hillenbrand}(2014)}]{Herczeg_14}
{Herczeg}, G.~J., \& {Hillenbrand}, L.~A. 2014, \apj, 786, 97

\bibitem[{{Hobbs} {et~al.}(2009){Hobbs}, {York}, {Thorburn}, {Snow}, {Bishof},
  {Friedman}, {McCall}, {Oka}, {Rachford}, {Sonnentrucker}, \&
  {Welty}}]{Hobbs_09}
{Hobbs}, L.~M., {York}, D.~G., {Thorburn}, J.~A., {et~al.} 2009, \apj, 705, 32

\bibitem[{Hunter(2007)}]{Hunter_07}
Hunter, J.~D. 2007, Computing In Science \& Engineering, 9, 90

\bibitem[{{Hur} {et~al.}(2015){Hur}, {Park}, {Sung}, {Bessell}, {Lim}, {Chun},
  \& {Sohn}}]{Hur_15}
{Hur}, H., {Park}, B.-G., {Sung}, H., {et~al.} 2015, \mnras, 446, 3797

\bibitem[{{Kaler}(2011)}]{Kaler_11}
{Kaler}, J.~B. 2011, {Stars and their Spectra} (Cambridge University Press)

\bibitem[{{Kamann} {et~al.}(2013){Kamann}, {Wisotzki}, \& {Roth}}]{Kamann_13}
{Kamann}, S., {Wisotzki}, L., \& {Roth}, M.~M. 2013, \aap, 549, A71

\bibitem[{{Kamann} {et~al.}(2016){Kamann}, {Husser}, {Brinchmann}, {Emsellem},
  {Weilbacher}, {Wisotzki}, {Wendt}, {Krajnovi{\'c}}, {Roth}, {Bacon}, \&
  {Dreizler}}]{Kamann_16}
{Kamann}, S., {Husser}, T.-O., {Brinchmann}, J., {et~al.} 2016, \aap, 588, A149

\bibitem[{{Kamann} {et~al.}(2018){Kamann}, {Husser}, {Dreizler}, {Emsellem},
  {Weilbacher}, {Martens}, {Bacon}, {den Brok}, {Giesers}, {Krajnovi{\'c}},
  {Roth}, {Wendt}, \& {Wisotzki}}]{Kamann_18}
{Kamann}, S., {Husser}, T.-O., {Dreizler}, S., {et~al.} 2018, \mnras, 473, 5591

\bibitem[{{Kiminki} \& {Smith}(2018)}]{Kimiki_18}
{Kiminki}, M.~M., \& {Smith}, N. 2018, \mnras, 477, 2068

\bibitem[{{Kobulnicky} {et~al.}(2012){Kobulnicky}, {Lundquist},
  {Bhattacharjee}, \& {Kerton}}]{Kobulnicky_12}
{Kobulnicky}, H.~A., {Lundquist}, M.~J., {Bhattacharjee}, A., \& {Kerton},
  C.~R. 2012, \aj, 143, 71

\bibitem[{{Kounkel} {et~al.}(2018){Kounkel}, {Covey}, {Su{\'a}rez},
  {Rom{\'a}n-Z{\'u}{\~n}iga}, {Hernandez}, {Stassun}, {Jaehnig}, {Feigelson},
  {Pe{\~n}a Ram{\'{\i}}rez}, {Roman-Lopes}, {Da Rio}, {Stringfellow}, {Kim},
  {Borissova}, {Fern{\'a}ndez-Trincado}, {Burgasser},
  {Garc{\'{\i}}a-Hern{\'a}ndez}, {Zamora}, {Pan}, \&
  {Nitschelm}}]{Kounkel_2018}
{Kounkel}, M., {Covey}, K., {Su{\'a}rez}, G., {et~al.} 2018, \aj, 156, 84

\bibitem[{{Kruijssen}(2015)}]{Kruijssen_15}
{Kruijssen}, J.~M.~D. 2015, ArXiv e-prints, arXiv:1509.02912

\bibitem[{{Lada} {et~al.}(1984){Lada}, {Margulis}, \& {Dearborn}}]{Lada_84a}
{Lada}, C.~J., {Margulis}, M., \& {Dearborn}, D. 1984, \apj, 285, 141

\bibitem[{{Laidler et al.}(2005)}]{Synphot}
{Laidler et al.} 2005, {Synphot Users's Guide}, Vol. Version 5.0 (Baltimore:
  STScI)

\bibitem[{{Martins} {et~al.}(2005){Martins}, {Schaerer}, \&
  {Hillier}}]{Martins_05}
{Martins}, F., {Schaerer}, D., \& {Hillier}, D.~J. 2005, \aap, 436, 1049

\bibitem[{{Massey}(2003)}]{Massey_03}
{Massey}, P. 2003, \araa, 41, 15

\bibitem[{{McLeod} {et~al.}(2015){McLeod}, {Dale}, {Ginsburg}, {Ercolano},
  {Gritschneder}, {Ramsay}, \& {Testi}}]{McLeod_15}
{McLeod}, A.~F., {Dale}, J.~E., {Ginsburg}, A., {et~al.} 2015, \mnras, 450,
  1057

\bibitem[{{McLeod} {et~al.}(2016){McLeod}, {Gritschneder}, {Dale}, {Ginsburg},
  {Klaassen}, {Mottram}, {Preibisch}, {Ramsay}, {Reiter}, \&
  {Testi}}]{McLeod_16}
{McLeod}, A.~F., {Gritschneder}, M., {Dale}, J.~E., {et~al.} 2016, \mnras, 462,
  3537

\bibitem[{{Moffat} {et~al.}(1991){Moffat}, {Shara}, \& {Potter}}]{Moffat_91}
{Moffat}, A.~F.~J., {Shara}, M.~M., \& {Potter}, M. 1991, \aj, 102, 642

\bibitem[{{Munari} \& {Tomasella}(1999)}]{Munari_99}
{Munari}, U., \& {Tomasella}, L. 1999, \aaps, 137, 521

\bibitem[{{Nigra} {et~al.}(2008){Nigra}, {Gallagher}, {Smith},
  {Stanimirovi{\'c}}, {Nota}, \& {Sabbi}}]{Nigra_08}
{Nigra}, L., {Gallagher}, J.~S., {Smith}, L.~J., {et~al.} 2008, \pasp, 120, 972

\bibitem[{{Pang} {et~al.}(2013){Pang}, {Grebel}, {Allison}, {Goodwin},
  {Altmann}, {Harbeck}, {Moffat}, \& {Drissen}}]{Pang_13}
{Pang}, X., {Grebel}, E.~K., {Allison}, R.~J., {et~al.} 2013, \apj, 764, 73

\bibitem[{{Parker} {et~al.}(2014){Parker}, {Wright}, {Goodwin}, \&
  {Meyer}}]{Parker_14}
{Parker}, R.~J., {Wright}, N.~J., {Goodwin}, S.~P., \& {Meyer}, M.~R. 2014,
  \mnras, 438, 620

\bibitem[{{Paxton} {et~al.}(2011){Paxton}, {Bildsten}, {Dotter}, {Herwig},
  {Lesaffre}, \& {Timmes}}]{Paxton_11}
{Paxton}, B., {Bildsten}, L., {Dotter}, A., {et~al.} 2011, \apjs, 192, 3

\bibitem[{{Paxton} {et~al.}(2013){Paxton}, {Cantiello}, {Arras}, {Bildsten},
  {Brown}, {Dotter}, {Mankovich}, {Montgomery}, {Stello}, {Timmes}, \&
  {Townsend}}]{Paxton_13}
{Paxton}, B., {Cantiello}, M., {Arras}, P., {et~al.} 2013, \apjs, 208, 4

\bibitem[{{Paxton} {et~al.}(2015){Paxton}, {Marchant}, {Schwab}, {Bauer},
  {Bildsten}, {Cantiello}, {Dessart}, {Farmer}, {Hu}, {Langer}, {Townsend},
  {Townsley}, \& {Timmes}}]{Paxton_15}
{Paxton}, B., {Marchant}, P., {Schwab}, J., {et~al.} 2015, \apjs, 220, 15

\bibitem[{{Rauw} {et~al.}(2007){Rauw}, {Manfroid}, {Gosset}, {Naz{\'e}},
  {Sana}, {De Becker}, {Foellmi}, \& {Moffat}}]{Rauw_07}
{Rauw}, G., {Manfroid}, J., {Gosset}, E., {et~al.} 2007, \aap, 463, 981

\bibitem[{{Rauw} {et~al.}(2011){Rauw}, {Sana}, \& {Naz{\'e}}}]{Rauw_11}
{Rauw}, G., {Sana}, H., \& {Naz{\'e}}, Y. 2011, \aap, 535, A40

\bibitem[{{Rochau} {et~al.}(2010){Rochau}, {Brandner}, {Stolte}, {Gennaro},
  {Gouliermis}, {Da Rio}, {Dzyurkevich}, \& {Henning}}]{Rochau_10}
{Rochau}, B., {Brandner}, W., {Stolte}, A., {et~al.} 2010, \apjl, 716, L90

\bibitem[{{Rodgers} {et~al.}(1960){Rodgers}, {Campbell}, \&
  {Whiteoak}}]{Rodgers_60}
{Rodgers}, A.~W., {Campbell}, C.~T., \& {Whiteoak}, J.~B. 1960, \mnras, 121,
  103

\bibitem[{{Sabbi} {et~al.}(2008){Sabbi}, {Sirianni}, {Nota}, {Tosi},
  {Gallagher}, {Smith}, {Angeretti}, {Meixner}, {Oey}, {Walterbos}, \&
  {Pasquali}}]{Sabbi_08}
{Sabbi}, E., {Sirianni}, M., {Nota}, A., {et~al.} 2008, \aj, 135, 173

\bibitem[{{Salpeter}(1955)}]{Salpeter_55}
{Salpeter}, E.~E. 1955, \apj, 121, 161

\bibitem[{{Smith} \& {Brooks}(2008)}]{Smith_08b}
{Smith}, N., \& {Brooks}, K.~J. 2008, {The Carina Nebula: A Laboratory for
  Feedback and Triggered Star Formation}, Vol.~8 (Astronomical Society of the
  Pacific), 138

\bibitem[{{Sota} {et~al.}(2014){Sota}, {Ma{\'{\i}}z Apell{\'a}niz}, {Morrell},
  {Barb{\'a}}, {Walborn}, {Gamen}, {Arias}, \& {Alfaro}}]{Sota_14}
{Sota}, A., {Ma{\'{\i}}z Apell{\'a}niz}, J., {Morrell}, N.~I., {et~al.} 2014,
  \apjs, 211, 10

\bibitem[{{Sota} {et~al.}(2011){Sota}, {Ma{\'{\i}}z Apell{\'a}niz}, {Walborn},
  {Alfaro}, {Barb{\'a}}, {Morrell}, {Gamen}, \& {Arias}}]{Sota_11}
{Sota}, A., {Ma{\'{\i}}z Apell{\'a}niz}, J., {Walborn}, N.~R., {et~al.} 2011,
  \apjs, 193, 24

\bibitem[{{The Astropy Collaboration} {et~al.}(2018){The Astropy
  Collaboration}, {Price-Whelan}, {Sip{\H o}cz}, {G{\"u}nther}, {Lim},
  {Crawford}, {Conseil}, {Shupe}, {Craig}, {Dencheva}, {Ginsburg},
  {VanderPlas}, {Bradley}, {P{\'e}rez-Su{\'a}rez}, {de Val-Borro}, {Aldcroft},
  {Cruz}, {Robitaille}, {Tollerud}, {Ardelean}, {Babej}, {Bachetti}, {Bakanov},
  {Bamford}, {Barentsen}, {Barmby}, {Baumbach}, {Berry}, {Biscani}, {Boquien},
  {Bostroem}, {Bouma}, {Brammer}, {Bray}, {Breytenbach}, {Buddelmeijer},
  {Burke}, {Calderone}, {Cano Rodr{\'{\i}}guez}, {Cara}, {Cardoso},
  {Cheedella}, {Copin}, {Crichton}, {D{\'A}vella}, {Deil}, {Depagne},
  {Dietrich}, {Donath}, {Droettboom}, {Earl}, {Erben}, {Fabbro}, {Ferreira},
  {Finethy}, {Fox}, {Garrison}, {Gibbons}, {Goldstein}, {Gommers}, {Greco},
  {Greenfield}, {Groener}, {Grollier}, {Hagen}, {Hirst}, {Homeier}, {Horton},
  {Hosseinzadeh}, {Hu}, {Hunkeler}, {Ivezi{\'c}}, {Jain}, {Jenness}, {Kanarek},
  {Kendrew}, {Kern}, {Kerzendorf}, {Khvalko}, {King}, {Kirkby}, {Kulkarni},
  {Kumar}, {Lee}, {Lenz}, {Littlefair}, {Ma}, {Macleod}, {Mastropietro},
  {McCully}, {Montagnac}, {Morris}, {Mueller}, {Mumford}, {Muna}, {Murphy},
  {Nelson}, {Nguyen}, {Ninan}, {N{\"o}the}, {Ogaz}, {Oh}, {Parejko}, {Parley},
  {Pascual}, {Patil}, {Patil}, {Plunkett}, {Prochaska}, {Rastogi}, {Reddy
  Janga}, {Sabater}, {Sakurikar}, {Seifert}, {Sherbert}, {Sherwood-Taylor},
  {Shih}, {Sick}, {Silbiger}, {Singanamalla}, {Singer}, {Sladen}, {Sooley},
  {Sornarajah}, {Streicher}, {Teuben}, {Thomas}, {Tremblay}, {Turner},
  {Terr{\'o}n}, {van Kerkwijk}, {de la Vega}, {Watkins}, {Weaver}, {Whitmore},
  {Woillez}, \& {Zabalza}}]{Astropy_18}
{The Astropy Collaboration}, {Price-Whelan}, A.~M., {Sip{\H o}cz}, B.~M.,
  {et~al.} 2018, ArXiv e-prints, arXiv:1801.02634

\bibitem[{{Underhill} {et~al.}(1979){Underhill}, {Divan}, {Prevot-Burnichon},
  \& {Doazan}}]{Underhill_79}
{Underhill}, A.~B., {Divan}, L., {Prevot-Burnichon}, M.-L., \& {Doazan}, V.
  1979, \mnras, 189, 601

\bibitem[{{Vargas {\'A}lvarez} {et~al.}(2013){Vargas {\'A}lvarez},
  {Kobulnicky}, {Bradley}, {Kannappan}, {Norris}, {Cool}, \&
  {Miller}}]{Vargas_Alvarez_13}
{Vargas {\'A}lvarez}, C.~A., {Kobulnicky}, H.~A., {Bradley}, D.~R., {et~al.}
  2013, \aj, 145, 125

\bibitem[{{Voggel} {et~al.}(2016){Voggel}, {Hilker}, {Baumgardt}, {Collins},
  {Grebel}, {Husemann}, {Richtler}, \& {Frank}}]{Voggel_16}
{Voggel}, K., {Hilker}, M., {Baumgardt}, H., {et~al.} 2016, \mnras, 460, 3384

\bibitem[{{Walborn} \& {Fitzpatrick}(1990)}]{Walborn_90}
{Walborn}, N.~R., \& {Fitzpatrick}, E.~L. 1990, \pasp, 102, 379

\bibitem[{{Weilbacher} {et~al.}(2012){Weilbacher}, {Streicher}, {Urrutia},
  {Jarno}, {P{\'e}contal-Rousset}, {Bacon}, \& {B{\"o}hm}}]{Weilbacher_12}
{Weilbacher}, P.~M., {Streicher}, O., {Urrutia}, T., {et~al.} 2012, in
  \procspie, Vol. 8451, Software and Cyberinfrastructure for Astronomy II,
  84510B

\bibitem[{{Weilbacher} {et~al.}(2014){Weilbacher}, {Streicher}, {Urrutia},
  {P{\'e}contal-Rousset}, {Jarno}, \& {Bacon}}]{Weilbacher_14}
{Weilbacher}, P.~M., {Streicher}, O., {Urrutia}, T., {et~al.} 2014, in
  Astronomical Society of the Pacific Conference Series, Vol. 485, Astronomical
  Data Analysis Software and Systems XXIII, ed. N.~{Manset} \& P.~{Forshay},
  451

\bibitem[{{Westerlund}(1961)}]{Westerlund_61}
{Westerlund}, B. 1961, Arkiv for Astronomi, 2, 419

\bibitem[{{Zeidler} \& {et}({2018, in prep.})}]{Zeidler_18b}
{Zeidler}, P., \& {et}, a.~l. {2018, in prep.}, \aj

\bibitem[{{Zeidler} {et~al.}(2016){Zeidler}, {Grebel}, {Nota}, {Sabbi},
  {Pasquali}, {Tosi}, {Bonanos}, \& {Christian}}]{Zeidler_16b}
{Zeidler}, P., {Grebel}, E.~K., {Nota}, A., {et~al.} 2016, \aj, 152, 84

\bibitem[{{Zeidler} {et~al.}(2017){Zeidler}, {Nota}, {Grebel}, {Sabbi},
  {Pasquali}, {Tosi}, \& {Christian}}]{Zeidler_17}
{Zeidler}, P., {Nota}, A., {Grebel}, E.~K., {et~al.} 2017, \aj, 153, 122

\bibitem[{{Zeidler} {et~al.}(2015){Zeidler}, {Sabbi}, {Nota}, {Grebel}, {Tosi},
  {Bonanos}, {Pasquali}, {Christian}, {de Mink}, \& {Ubeda}}]{Zeidler_15}
{Zeidler}, P., {Sabbi}, E., {Nota}, A., {et~al.} 2015, \aj, 150, 78

\end{thebibliography}

\appendix
\section{The properties of the stars used in this work}

In Tab.~\ref{tab:spec_stars} we present all stars and their properties used in the analyses of this paper.

\startlongtable
\begin{deluxetable*}{rrrccccrrcl}
	\tablecaption{The analyzed stars  \label{tab:spec_stars}}
	\tabletypesize{\scriptsize}
	\tablewidth{0pt}
	\tablehead{
		\multicolumn{1}{c}{ID} &\multicolumn{1}{c}{R.A.} & \multicolumn{1}{c}{Dec.} &  \multicolumn{1}{c}{$F555W$} &  \multicolumn{1}{c}{$F814W$}  & \multicolumn{1}{c}{$F160W$} &  \multicolumn{1}{c}{$F814W_{\rm spec}$} & \multicolumn{1}{c}{rv} & \multicolumn{1}{c}{ $\sigma_{\rm rv}$} & \multicolumn{1}{c}{ ${\rm N}_{\rm lines}$}  & \multicolumn{1}{c}{spectral type} \\		
		\multicolumn{1}{c}{ } &\multicolumn{2}{c}{(J2000)}  &  \multicolumn{4}{c}{[mag]}  & \multicolumn{2}{c}{[km/s]}  & \multicolumn{2}{c}{ }
	}
	\tablecolumns{11}
	\startdata
	\cutinhead{Northern Bubble stars}
6660 & $10^\mathrm{h}24^\mathrm{m}00.46^\mathrm{s}$ & $-57^\circ44{}^\prime37.43{}^{\prime\prime}$ & 17.459 & 15.157 & 12.751 & 14.952 & -1.168 & 3.273 & 3 &  B4.5\tablenotemark{a}\\
6687 & $10^\mathrm{h}24^\mathrm{m}00.49^\mathrm{s}$ & $-57^\circ44{}^\prime44.45{}^{\prime\prime}$ & 15.711 & 13.644 & --- & 13.409 & 29.302 & 4.852 & 1 &  B1.5\tablenotemark{b}\\
7057 & $10^\mathrm{h}24^\mathrm{m}00.85^\mathrm{s}$ & $-57^\circ44{}^\prime27.17{}^{\prime\prime}$ & 18.065 & 15.883 & 13.646 & 15.634 & 24.906 & 2.029 & 3 &  B4\\
7510 & $10^\mathrm{h}24^\mathrm{m}01.26^\mathrm{s}$ & $-57^\circ44{}^\prime23.01{}^{\prime\prime}$ & 16.496 & 14.282 & 12.150 & 14.026 & 15.132 & 1.713 & 2 &  B0\\
8586 & $10^\mathrm{h}24^\mathrm{m}02.22^\mathrm{s}$ & $-57^\circ44{}^\prime35.48{}^{\prime\prime}$ & 18.030 & 15.733 & 13.367 & 15.318 & -2.446 & 4.840 & 2 & --- \\
8768 & $10^\mathrm{h}24^\mathrm{m}02.39^\mathrm{s}$ & $-57^\circ44{}^\prime34.80{}^{\prime\prime}$ & 15.512 & 13.502 & --- & 13.163 & 9.851 & 2.571 & 2 &  O7.5\\
8806 & $10^\mathrm{h}24^\mathrm{m}02.43^\mathrm{s}$ & $-57^\circ44{}^\prime36.09{}^{\prime\prime}$ & 13.076 & --- & --- & 10.794 & 2.611 & 1.038 & 2 &  \textit{O5V-III((f))}\tablenotemark{c}\\
9186 & $10^\mathrm{h}24^\mathrm{m}02.78^\mathrm{s}$ & $-57^\circ44{}^\prime37.64{}^{\prime\prime}$ & 16.508 & 14.579 & 12.769 & 14.365 & 8.184 & 1.138 & 2 &  B1.5\\
9529 & $10^\mathrm{h}24^\mathrm{m}03.09^\mathrm{s}$ & $-57^\circ44{}^\prime17.48{}^{\prime\prime}$ & --- & 14.480 & 12.730 & 14.480 & 58.227 & 2.959 & 3  &  B5\\
10198 & $10^\mathrm{h}24^\mathrm{m}03.77^\mathrm{s}$ & $-57^\circ44{}^\prime39.79{}^{\prime\prime}$ & 15.414 & 13.328 & --- & 13.091 & 29.056 & 1.156 & 4 &  O9.5\tablenotemark{d}\\
10225 & $10^\mathrm{h}24^\mathrm{m}03.80^\mathrm{s}$ & $-57^\circ44{}^\prime34.92{}^{\prime\prime}$ & 18.539 & 15.600 & 13.084 & 15.154 & 7.745 & 1.112 & 2  &  G5\\
10372 & $10^\mathrm{h}24^\mathrm{m}03.97^\mathrm{s}$ & $-57^\circ44{}^\prime36.59{}^{\prime\prime}$ & 15.615 & 13.585 & --- & 13.399 & 27.453 & 3.622 & 3 &  O8.5\\
10759 & $10^\mathrm{h}24^\mathrm{m}04.40^\mathrm{s}$ & $-57^\circ44{}^\prime36.17{}^{\prime\prime}$ & 17.268 & 14.674 & 12.584 & 14.375 & 22.216 & 2.598 & 2 &  B4.5\\
11126 & $10^\mathrm{h}24^\mathrm{m}04.83^\mathrm{s}$ & $-57^\circ44{}^\prime11.42{}^{\prime\prime}$ & 19.446 & 17.167 & 14.984 & --- & 2.857 & 1.927 & 2 & --- \\
11414 & $10^\mathrm{h}24^\mathrm{m}05.17^\mathrm{s}$ & $-57^\circ44{}^\prime36.65{}^{\prime\prime}$ & 19.148 & 16.380 & 13.733 & --- & 0.025 & 3.829 & 2 & --- \\
12258 & $10^\mathrm{h}24^\mathrm{m}06.21^\mathrm{s}$ & $-57^\circ44{}^\prime42.96{}^{\prime\prime}$ & 17.943 & 15.439 & 12.663 & --- & -48.172 & 9.338 & 2 & --- \\
\cutinhead{cluster members in the remaining Wd2 cluster}
2651 & $10^\mathrm{h}23^\mathrm{m}55.16^\mathrm{s}$ & $-57^\circ45{}^\prime26.88{}^{\prime\prime}$ & --- & --- & --- & --- & 25.505 & 8.789 & 2 &  --- \\
3077 & $10^\mathrm{h}23^\mathrm{m}55.88^\mathrm{s}$ & $-57^\circ44{}^\prime24.01{}^{\prime\prime}$ & 18.441 & 16.007 & --- & --- & 6.693 & 6.642 & 2 &  --- \\
3904 & $10^\mathrm{h}23^\mathrm{m}57.23^\mathrm{s}$ & $-57^\circ44{}^\prime37.91{}^{\prime\prime}$ & 16.358 & 13.918 & --- & --- & 10.984 & 3.224 & 4 &  --- \\
3934 & $10^\mathrm{h}23^\mathrm{m}57.26^\mathrm{s}$ & $-57^\circ44{}^\prime06.98{}^{\prime\prime}$ & --- & 25.212 & 19.112 & --- & 14.614 & 4.919 & 2 &  --- \\
4221 & $10^\mathrm{h}23^\mathrm{m}57.68^\mathrm{s}$ & $-57^\circ45{}^\prime26.68{}^{\prime\prime}$ & 20.528 & 17.089 & 13.612 & --- & -31.711 & 3.995 & 3 &  --- \\
4246 & $10^\mathrm{h}23^\mathrm{m}57.71^\mathrm{s}$ & $-57^\circ45{}^\prime34.24{}^{\prime\prime}$ & 16.959 & 14.649 & 12.387 & --- & 20.387 & 4.927 & 2 &  --- \\
4821 & $10^\mathrm{h}23^\mathrm{m}58.43^\mathrm{s}$ & $-57^\circ45{}^\prime12.95{}^{\prime\prime}$ & 16.287 & 13.770 & --- & --- & 12.586 & 2.737 & 4 &  --- \\
4948 & $10^\mathrm{h}23^\mathrm{m}58.58^\mathrm{s}$ & $-57^\circ45{}^\prime05.40{}^{\prime\prime}$ & 19.370 & 16.829 & 13.298 & --- & 5.376 & 10.178 & 1 &  --- \\
5476 & $10^\mathrm{h}23^\mathrm{m}59.24^\mathrm{s}$ & $-57^\circ45{}^\prime39.73{}^{\prime\prime}$ & 18.403 & 15.926 & 13.509 & 15.593 & 23.599 & 5.311 & 2 &  --- \\
5505 & $10^\mathrm{h}23^\mathrm{m}59.26^\mathrm{s}$ & $-57^\circ45{}^\prime05.67{}^{\prime\prime}$ & 19.043 & 16.462 & 13.746 & --- & -3.682 & 2.850 & 2 &  --- \\
5773 & $10^\mathrm{h}23^\mathrm{m}59.54^\mathrm{s}$ & $-57^\circ45{}^\prime22.49{}^{\prime\prime}$ & 13.042 & 11.800 & --- & 11.534 & 18.179 & 0.168 & 2 &  --- \\
5870 & $10^\mathrm{h}23^\mathrm{m}59.66^\mathrm{s}$ & $-57^\circ45{}^\prime06.48{}^{\prime\prime}$ & 17.368 & 15.182 & 13.096 & 14.903 & 10.906 & 2.301 & 2 &  --- \\
6019 & $10^\mathrm{h}23^\mathrm{m}59.82^\mathrm{s}$ & $-57^\circ45{}^\prime15.58{}^{\prime\prime}$ & 17.863 & 15.680 & 13.601 & 15.437 & 7.387 & 3.984 & 2 &  --- \\
6342 & $10^\mathrm{h}24^\mathrm{m}00.14^\mathrm{s}$ & $-57^\circ45{}^\prime29.76{}^{\prime\prime}$ & 18.145 & 15.519 & 12.962 & 15.277 & 3.832 & 4.970 & 2 &  --- \\
6391 & $10^\mathrm{h}24^\mathrm{m}00.18^\mathrm{s}$ & $-57^\circ45{}^\prime32.53{}^{\prime\prime}$ & 16.102 & 13.861 & 11.692 & 13.570 & 12.539 & 1.961 & 4 &  --- \\
6528 & $10^\mathrm{h}24^\mathrm{m}00.34^\mathrm{s}$ & $-57^\circ45{}^\prime42.62{}^{\prime\prime}$ & 15.880 & 13.510 & --- & 13.171 & 37.477 & 5.312 & 1 &  --- \\
7620 & $10^\mathrm{h}24^\mathrm{m}01.38^\mathrm{s}$ & $-57^\circ45{}^\prime29.58{}^{\prime\prime}$ & 14.085 & 12.045 & --- & 11.705 & 4.209 & 2.534 & 2 &  --- \\
8174 & $10^\mathrm{h}24^\mathrm{m}01.85^\mathrm{s}$ & $-57^\circ45{}^\prime31.17{}^{\prime\prime}$ & 15.917 & 13.758 & --- & 13.576 & 9.065 & 2.807 & 2 &  --- \\
8728 & $10^\mathrm{h}24^\mathrm{m}02.35^\mathrm{s}$ & $-57^\circ45{}^\prime30.56{}^{\prime\prime}$ & 13.942 & --- & --- & 11.181 & 29.419 & 1.512 & 1 &  --- \\
9013 & $10^\mathrm{h}24^\mathrm{m}02.62^\mathrm{s}$ & $-57^\circ45{}^\prime37.43{}^{\prime\prime}$ & 16.143 & 13.965 & 12.007 & 13.791 & -13.374 & 3.378 & 2 &  --- \\
10048 & $10^\mathrm{h}24^\mathrm{m}03.59^\mathrm{s}$ & $-57^\circ45{}^\prime27.08{}^{\prime\prime}$ & 16.809 & 14.551 & 12.461 & 14.285 & 7.546 & 3.194 & 2 &  --- \\
11178 & $10^\mathrm{h}24^\mathrm{m}04.89^\mathrm{s}$ & $-57^\circ45{}^\prime28.35{}^{\prime\prime}$ & 14.607 & 12.254 & --- & 11.963 & -8.204 & 1.326 & 3 &  --- \\
12154 & $10^\mathrm{h}24^\mathrm{m}06.09^\mathrm{s}$ & $-57^\circ45{}^\prime57.65{}^{\prime\prime}$ & 18.008 & 15.342 & 12.913 & 15.567 & 33.474 & 3.056 & 2 &  --- \\
13585 & $10^\mathrm{h}24^\mathrm{m}07.91^\mathrm{s}$ & $-57^\circ44{}^\prime23.46{}^{\prime\prime}$ & 17.395 & 15.451 & 13.702 & --- & 5.915 & 11.727 & 1 &  --- \\
14138 & $10^\mathrm{h}24^\mathrm{m}08.68^\mathrm{s}$ & $-57^\circ46{}^\prime02.42{}^{\prime\prime}$ & 18.165 & 15.828 & 13.380 & 16.076 & 13.768 & 2.527 & 2 &  --- \\
14161 & $10^\mathrm{h}24^\mathrm{m}08.74^\mathrm{s}$ & $-57^\circ44{}^\prime43.97{}^{\prime\prime}$ & 19.653 & 17.121 & 14.563 & --- & -6.481 & 2.515 & 5 &  --- \\
14213 & $10^\mathrm{h}24^\mathrm{m}08.82^\mathrm{s}$ & $-57^\circ44{}^\prime36.56{}^{\prime\prime}$ & 16.666 & 14.834 & --- & --- & 19.159 & 0.717 & 5 &  --- \\
15613 & $10^\mathrm{h}24^\mathrm{m}10.69^\mathrm{s}$ & $-57^\circ46{}^\prime00.00{}^{\prime\prime}$ & 16.387 & 12.783 & --- & 12.983 & -11.379 & 0.171 & 3 &  --- \\
\cutinhead{Field stars in the Northern Bubble}
6446 & $10^\mathrm{h}24^\mathrm{m}00.25^\mathrm{s}$ & $-57^\circ44{}^\prime07.96{}^{\prime\prime}$ & --- & --- & 12.690 & 13.230 & 4.480 & 0.212 & 6 &  F7V\\
6542 & $10^\mathrm{h}24^\mathrm{m}00.35^\mathrm{s}$ & $-57^\circ44{}^\prime17.50{}^{\prime\prime}$ & 18.035 & 17.051 & 15.686 & 16.655 & 16.792 & 1.254 & 4 &  G8III-IV\\
6717 & $10^\mathrm{h}24^\mathrm{m}00.51^\mathrm{s}$ & $-57^\circ44{}^\prime26.49{}^{\prime\prime}$ & 17.171 & 16.005 & 14.879 & 15.843 & -0.896 & 1.212 & 2 &  G5V\tablenotemark{a}\\
7830 & $10^\mathrm{h}24^\mathrm{m}01.55^\mathrm{s}$ & $-57^\circ44{}^\prime05.68{}^{\prime\prime}$ & --- & --- & 12.285 & 12.514 & 17.746 & 0.353 & 3 &  A9V\\
10615 & $10^\mathrm{h}24^\mathrm{m}04.26^\mathrm{s}$ & $-57^\circ44{}^\prime22.92{}^{\prime\prime}$ & 16.344 & 15.387 & 14.448 & 15.216 & 15.557 & 0.448 & 4 &   G8IV var\\
10733 & $10^\mathrm{h}24^\mathrm{m}04.37^\mathrm{s}$ & $-57^\circ44{}^\prime28.51{}^{\prime\prime}$ & 17.884 & 16.938 & 15.362 & 16.390 & 43.358 & 1.496 & 2 &  G9III\\
11065 & $10^\mathrm{h}24^\mathrm{m}04.75^\mathrm{s}$ & $-57^\circ44{}^\prime32.95{}^{\prime\prime}$ & 16.469 & 15.587 & 14.718 & 15.409 & 3.263 & 0.610 & 4 &  --- \\
11549 & $10^\mathrm{h}24^\mathrm{m}05.32^\mathrm{s}$ & $-57^\circ44{}^\prime44.24{}^{\prime\prime}$ & 19.082 & 17.037 & 15.093 & --- & 9.966 & 2.315 & 3 &  --- \\
\cutinhead{Field stars in the remaining Wd2 cluster}
5753 & $10^\mathrm{h}23^\mathrm{m}59.52^\mathrm{s}$ & $-57^\circ46{}^\prime42.55{}^{\prime\prime}$ & 15.893 & 15.036 & 14.232 & 14.940 & -30.347 & 0.476 & 5 &  --- \\
6973 & $10^\mathrm{h}24^\mathrm{m}00.75^\mathrm{s}$ & $-57^\circ46{}^\prime29.58{}^{\prime\prime}$ & 14.291 & 13.506 & 12.721 & 13.397 & 14.301 & 0.268 & 5 &  --- \\
7109 & $10^\mathrm{h}24^\mathrm{m}00.90^\mathrm{s}$ & $-57^\circ46{}^\prime10.69{}^{\prime\prime}$ & 16.553 & 15.621 & 14.684 & 15.503 & 13.477 & 1.807 & 2 &  --- \\
7282 & $10^\mathrm{h}24^\mathrm{m}01.06^\mathrm{s}$ & $-57^\circ46{}^\prime18.63{}^{\prime\prime}$ & 15.735 & 14.808 & 13.899 & 14.717 & -21.500 & 0.491 & 2 &  --- \\
8584 & $10^\mathrm{h}24^\mathrm{m}02.21^\mathrm{s}$ & $-57^\circ45{}^\prime02.65{}^{\prime\prime}$ & 13.869 & 13.032 & 12.271 & 12.837 & -6.442 & 0.209 & 6 &  --- \\
9577 & $10^\mathrm{h}24^\mathrm{m}03.14^\mathrm{s}$ & $-57^\circ45{}^\prime21.12{}^{\prime\prime}$ & 17.017 & 15.890 & 14.776 & 15.645 & -43.813 & 0.796 & 4 &  --- \\
9709 & $10^\mathrm{h}24^\mathrm{m}03.26^\mathrm{s}$ & $-57^\circ45{}^\prime50.83{}^{\prime\prime}$ & 17.389 & 16.356 & 15.334 & 16.221 & -29.858 & 0.695 & 5 &  --- \\
9849 & $10^\mathrm{h}24^\mathrm{m}03.40^\mathrm{s}$ & $-57^\circ46{}^\prime09.64{}^{\prime\prime}$ & 15.065 & 14.308 & 13.600 & 14.214 & 25.433 & 0.455 & 3 &  --- \\
11134 & $10^\mathrm{h}24^\mathrm{m}04.84^\mathrm{s}$ & $-57^\circ45{}^\prime49.88{}^{\prime\prime}$ & 16.970 & 15.799 & 14.653 & 15.648 & 12.653 & 0.707 & 5 &  --- \\
12265 & $10^\mathrm{h}24^\mathrm{m}06.21^\mathrm{s}$ & $-57^\circ46{}^\prime10.31{}^{\prime\prime}$ & 16.048 & 15.198 & 14.475 & 15.500 & 23.228 & 0.629 & 5 &  --- \\
12428 & $10^\mathrm{h}24^\mathrm{m}06.43^\mathrm{s}$ & $-57^\circ44{}^\prime50.85{}^{\prime\prime}$ & 16.860 & 15.850 & 14.862 & 15.680 & 5.333 & 1.051 & 6 &  --- \\
12826 & $10^\mathrm{h}24^\mathrm{m}06.89^\mathrm{s}$ & $-57^\circ46{}^\prime27.31{}^{\prime\prime}$ & 16.404 & 15.400 & 14.481 & 15.708 & 5.067 & 0.639 & 5 &  --- \\
12987 & $10^\mathrm{h}24^\mathrm{m}07.10^\mathrm{s}$ & $-57^\circ46{}^\prime16.62{}^{\prime\prime}$ & 15.841 & 14.921 & 14.050 & 15.243 & -4.665 & 0.756 & 3 &  --- \\
13727 & $10^\mathrm{h}24^\mathrm{m}08.10^\mathrm{s}$ & $-57^\circ45{}^\prime40.33{}^{\prime\prime}$ & 15.778 & 14.939 & 14.115 & 15.101 & -10.300 & 1.037 & 3 &  --- \\
13850 & $10^\mathrm{h}24^\mathrm{m}08.28^\mathrm{s}$ & $-57^\circ44{}^\prime28.22{}^{\prime\prime}$ & 17.377 & 16.354 & 15.439 & --- & -6.961 & 1.277 & 3 &  --- \\
13986 & $10^\mathrm{h}24^\mathrm{m}08.47^\mathrm{s}$ & $-57^\circ44{}^\prime28.84{}^{\prime\prime}$ & 16.390 & 15.537 & 14.739 & 15.542 & -13.665 & 1.817 & 2 &  --- \\
14186 & $10^\mathrm{h}24^\mathrm{m}08.78^\mathrm{s}$ & $-57^\circ44{}^\prime47.70{}^{\prime\prime}$ & 17.598 & 15.938 & 14.405 & --- & 0.731 & 2.466 & 3 &  --- \\
16080 & $10^\mathrm{h}24^\mathrm{m}11.38^\mathrm{s}$ & $-57^\circ45{}^\prime52.16{}^{\prime\prime}$ & 16.788 & 15.820 & 14.923 & 16.006 & -1.286 & 0.795 & 2 &  --- \\
17656 & $10^\mathrm{h}24^\mathrm{m}14.19^\mathrm{s}$ & $-57^\circ44{}^\prime44.92{}^{\prime\prime}$ & 14.734 & 14.038 & 13.425 & 14.099 & -6.172 & 0.760 & 4 &  --- \\
19296 & $10^\mathrm{h}24^\mathrm{m}18.27^\mathrm{s}$ & $-57^\circ45{}^\prime26.47{}^{\prime\prime}$ & --- & --- & --- & --- & -2.767 & 1.572 & 2 &  --- \\
	\enddata
	\tablecomments{In this table we summarize the properties of the 72 stars used in this work. Column 1 shows the identifier of our HST photometric catalog. Columns 2 \& 3 give the stellar coordinates. Columns 4--6 show the HST photometry used in this analysis. Column 7 represents the $F814W$ magnitude based on the extracted spectrum\tablenotemark{e}. Columns 8 \& 9 show the measured RVs of the stars including their uncertainty.  Column 11 gives the number of lines used to measure the RV. Column 11 shows the spectral type derived in Sect.~\ref{sec:spec_class}. Spectral types written in italics were recovered from the literature.}
	\tablenotetext{a}{\citet{Vargas_Alvarez_13} determined an upper limit of O-B}	
	\tablenotetext{b}{This star was classified as B1V by \citet{Rauw_07} and \citet{Rauw_11}}
	\tablenotetext{c}{\citet{Rauw_07} classified this star as O5V-III((f)).}
	\tablenotetext{d}{This star was classified as O9.5V by \citet{Vargas_Alvarez_13}}
	\tablenotetext{e}{The magnitude based on the spectrum was calculated by folding the respective spectrum with the ACS $F814W$ response curve and zeropoint using \texttt{pysynphot} \citep{Synphot}.}
\end{deluxetable*}

\section{The individual spectral classification}
\label{sec:individual_ST}

\#6446: The bluest Pa line visible is Pa19 and the missing \ion{Si}{2}$\lambda\lambda6347,6371$ exclude the star from being a supergiant. These criteria and the best fitted \ion{Ca}{2}-triplet suggest the star being a F7V star.

\#6717: This star shows a weak Paschen series in comparison to the very pronounced \ion{Ca}{2}-triplet. The last visible Paschen line is Pa17, which indicates a luminosity class V and a spectral type of late A, F, or G-type. A large number of ionized and neutral metals are visible (e.g., \ion{Mg}{1}-triplet, \ion{S}{2}, and many iron lines). The visibility of a weak \ion{Ti}{1}$\lambda8435$ line, which appears to be equally strong to the \ion{Fe}{1}$\lambda8468$ line, favors a spectral type later than early G. The best fitting template for  \ion{Ca}{2}-triplet is a G5V star which is in agreement with the above argumentation. \citet{Vargas_Alvarez_13} argue for a spectral type of late O to early B based on weak \ion{He}{1} lines. The existence of the \ion{Ca}{2}-triplet rules out an OB-star. This discrepancy was probably caused by a low S/N of the spectrum used in the analysis by \citet{Vargas_Alvarez_13}.

\#7830: This star has broad Balmer lines and strong Pa-lines with the presence of the \ion{Ca}{2}-triplet, although the Pa lines are clearly dominating. In combination with a very strong \ion{O}{1}-triplet ($\lambda\lambda7772,7774,7775$) this appears to be a late A-type star. The last visible Pa line is Pa18 suggesting a luminosity class of II, IV, or V. The fact that it still shows the  \ion{Ca}{2}-triplet allows to use the spectral library \citep{Cenarro_01} for classification, leading to a spectral type of A9V in agreement with the above argumentation.

\#10615 and \#6542: The appearance of ionized metals (e.g., \ion{C}{3}$\lambda4647$), an almost absent Pa-series, narrow Balmer lines, and the appearance of \ion{Ti}{1}$\lambda8435$ suggest a late G to early K-type star. The best-fitting model for \#10615 is a G8IV star while for \#6542 it is a G8III-IV star. 

\#10733: Showing narrow Balmer lines suggest a spectral type later than A0. The presence of many neutral metals, the lack of the \ion{O}{1}$\lambda7774$ line, and the almost absent Pa-series suggests a spectral type later than F5. The existence of the \ion{Ti}{1}$\lambda8435$ line together with neutral iron in the \ion{Ca}{2}-triplet favors a later G type to early K type star. The best fitting template is a G9III star.

\end{document}